\documentclass[manuscript, nonacm]{acmart}

\usepackage{xcolor, tcolorbox, multirow, dcolumn, threeparttable, color, subfig, graphicx, mathrsfs, amsmath, bbding, multicol}

\AtBeginDocument{%
  \providecommand\BibTeX{{%
    \normalfont B\kern-0.5em{\scshape i\kern-0.25em b}\kern-0.8em\TeX}}}

\setcopyright{acmcopyright}
\copyrightyear{}
\acmYear{}
\acmDOI{}

\acmBooktitle{CHI '22: ACM CHI Conference on Human Factors in Computing Systems,
April 30-May 6, 2022, New Orleans, LA}
\acmPrice{15.00}
\acmISBN{978-1-4503-XXXX-X/18/06}

\begin{document}

\title[The Development and Prospect of Code Clone]{The Development and Prospect of Code Clone}

\author{Xunhui Zhang}
\email{zhangxunhui@nudt.edu.cn}
\affiliation{%
  \institution{National University of Defense Technology}
  \country{China}
}

\author{Tao Wang}
\email{taowang2005@nudt.edu.cn}
\affiliation{%
  \institution{National University of Defense Technology}
  \country{China}
}

\author{Yue Yu}
\email{yuyue@nudt.edu.cn}
\affiliation{%
  \institution{National University of Defense Technology}
  \country{China}
}

\author{Yanzhi Zhang}
\affiliation{%
  \institution{National University of Defense Technology}
  \country{China}
}
\email{zhangyanzhi19@nudt.edu.cn}

\author{Yan Zhong}
\affiliation{%
  \institution{National University of Defense Technology}
  \country{China}
}
\email{MillerEvan@163.com}

\author{Huaimin Wang}
\affiliation{%
  \institution{National University of Defense Technology}
  \country{China}
}
\email{whm\_w@163.com}

\renewcommand{\shortauthors}{Zhang et al.}

\begin{abstract}
  The application of code clone technology accelerates code search, improves code reuse efficiency, and assists in software quality assessment and code vulnerability detection. However, the application of code clones also introduces software quality issues and increases the cost of software maintenance. As an important research field in software engineering, code clone has been extensively explored and studied by researchers, and related studies on various sub-research fields have emerged, including code clone detection, code clone evolution, code clone analysis, etc. However, there lacks a comprehensive exploration of the entire field of code clone, as well as an analysis of the trend of each sub-research field. This paper collects related work of code clones in the past ten years. In summary, the contributions of this paper mainly include: (1) summarize and classify the sub-research fields of code clone, and explore the relative popularity and relation of these sub-research fields; (2) analyze the overall research trend of code clone and each sub-research field; (3) compare and analyze the difference between academy and industry regarding code clone research; (4) construct a network of researchers, and excavate the major contributors in code clone research field; (5) The list of popular conferences and journals was statistically analyzed. The popular research directions in the future include clone visualization, clone management, etc. For the clone detection technique, researchers can optimize the scalability and execution efficiency of the method, targeting particular clone detection tasks and contextual environments, or apply the technology to other related research fields continuously.
\end{abstract}




\maketitle

\section{Introduction}
Code clone refers to two or more identical or similar source code fragments in a codebase~\cite{1}. As an important research topic in software engineering, code clone enhances software development efficiency aids software quality assessment and software vulnerability discovery. However, code clone is a bad smell that brings defects, which increases software maintenance costs and leads to software quality degradation. It also introduces intellectual property protection issues. Therefore, both industry and academia have paid close attention to the code clone problem.
Many studies focus on code clones, including clone detection~\cite{4,5}, clone evolution~\cite{6,7}, clone visualization~\cite{8,9}, clone refactoring\cite{10,11}, etc. In response to the above research problems, many literature reviews have emerged. However, these reviews mainly focus on sub-research areas such as clone detection~\cite{12}, clone evolution~\cite{13}, clone visualization~\cite{14}. However, there is a lack of reviews to scientifically summarize the code clone sub-research areas for the whole code clone field and analyze each sub-research area's development trend. 
This paper collects and organizes the work in code clone in the past ten years to address the problem. We extract the basic information of the articles, including the date of publication, title, abstract, keywords, author and affiliation information, etc. Then we classify the articles into sub-research areas by card sorting and finally analyze the development of each sub-research area of code clone by statistical methods. We also construct an author cooperation network from the perspective of researchers, explore the research pattern of code cloning, find the researchers who have made outstanding research contributions to the research of code clone and their affiliations, and analyze the difference of researcher cooperation among countries.
The main contributions of this paper are as follows:
\begin{enumerate}
    \item We collected 1,294 papers related to code clone in the last decade (from 2011 to 2020);
    \item A manual classification of code clone related sub-research areas was conducted, and a public dataset was formed, containing information on topics and related articles of sub-research areas;
    \item The hotness analysis of code cloning sub-research areas was conducted to explore the changes of code clone as a whole and each sub-research area in the past ten years;
    \item We compared and analyzed the changes of industrial and academic interest in each sub-research area of code clone and the overall interest in the sub-research area;
    \item We constructed a network of collaborative relationships among code clone researchers, analyzed the characteristics of collaborative relationships, identified outstanding contributors and popular research institutions, and explored the differences of collaborative relationships in different countries.
\end{enumerate}
Section 2 of this paper introduces the research background, summarizes the systematic literature review articles related to code cloning and the differences between this paper and related works, and presents the research questions of this paper.
Section 3 introduces collecting, screening, and data extraction of clone-related articles based on a systematic literature review.
Section 4 describes the classification method of code clone sub-research areas based on card sorting.
Section 5 presents the results of classification of code clone sub-research areas, trends of each category, the attention of industry and academia, analysis of author collaboration networks, and analysis of popular journals and conferences.
Section 6 discusses the paper's conclusions and gives an outlook for future research in the field of code cloning.
Finally, Section 7 concludes the paper.

\section{Background}
\label{background}
Although there have been many surveys on code cloning, most have focused on a sub-research area. Table~\ref{table_1} summarizes the systematic literature review related to code cloning, comparing several aspects, including the time of publication, the number of articles included, and the year covered by the article. The ``basic information'' and the ``research point'', which refers to the issue in the field of code cloning that the article focuses on, are all part of ``code clone detection''.
For example, Ain et al.~\cite{15} focus on code clone detection methods, the related data, and intermediate data representations involved in the methods. Although there are related works that focus on sub-research area classification and popularity analysis~\cite{16,12,18,19,20,21}, these works either focus on a specific sub-research area, e.g., software similarity~\cite{16}, or the research area has broadened. The popularity may change after a long time of development~\cite{12,18,20,21}, Or there is a lack of analysis of the development trend of sub-research areas~\cite{19}. At the same time, we found that the classification of code cloning in the relevant studies was based on research experience. The classification results varied widely, lacking the classification of code cloning sub-research areas based on scientific research methods. Finally, all the related studies lacked the analysis of the popularity of attention to code cloning in practice.

\begin{table}[htbp]
    \begin{center}
        \caption{Collection of basic information and focus of code clone related systematic literature reviews}
        \label{table_1}
        \renewcommand\arraystretch{1.25}
        \scriptsize
        \begin{tabular}{m{1.5cm}|m{0.35cm}|m{0.35cm}|m{0.35cm}|m{0.35cm}|m{0.35cm}|m{0.35cm}|m{0.35cm}|m{0.35cm}|m{0.35cm}|m{0.35cm}|m{0.35cm}|m{0.35cm}|m{0.35cm}|m{0.35cm}|m{0.35cm}|m{0.35cm}|m{0.35cm}|m{0.35cm}|m{0.35cm}|m{0.35cm}}
        \hline
        \textbf{Item} & \textbf{\cite{16}} & \textbf{\cite{13}} & \textbf{\cite{12}} & \textbf{\cite{22}} & \textbf{\cite{15}} & \textbf{\cite{23}} & \textbf{\cite{24}} & \textbf{\cite{25}} & \textbf{\cite{26}} & \textbf{\cite{27}} & \textbf{\cite{28}} & \textbf{\cite{14}} & \textbf{\cite{18}} & \textbf{\cite{29}} & \textbf{\cite{19}} & \textbf{\cite{20}} & \textbf{\cite{30}} & \textbf{\cite{31}} & \textbf{\cite{32}} & \textbf{\cite{21}}\\
        \hline
        Date & 2020 & 2013 & 2013 & 2014 & 2019 & 2017 & 2016 & 2014 & 2020 & 2017 & 2014 & 2020 & 2012 & 2019 & 2020 & 2012 & 2019 & 2020 & 2014 & 2018 \\
        \hline
        Paper num & 136 & 30 & 213 & 7 & 54 & 61 & 30 & 20 & 198 & 177 & 65 & 68 & 262 & 32 & 27 & 220 & 54 & 97 & 39 & 575 \\
        \hline
        Time range & 2002-2019 & -2011 & 1997-2011 & -2012 & 2013-2018 & 1997-2016 & 1996-2015 & 2010-2014 & 2011-2017 & 2011- & -2014 & -2020 & 1994-2011 & 2001-2018 & 1998-2017 & 2007-2011 & 2013-2018 & 1998-2017 & -2014 & -2013 \\
        \hline
        Sub-research area & \Checkmark & & \Checkmark & & & & & & & & & & \Checkmark & & \Checkmark & \Checkmark & & & & \Checkmark \\
        \hline
        Data & \Checkmark & \Checkmark & & & \Checkmark & & & & & & & & & \Checkmark & & \Checkmark & & & & \\
        \hline
        Clone method & & \Checkmark & \Checkmark & \Checkmark & \Checkmark & & & \Checkmark & & \Checkmark & \Checkmark & & \Checkmark & \Checkmark & \Checkmark & \Checkmark & \Checkmark & & & \Checkmark \\
        \hline
        Code representation & & & & & \Checkmark & & & & & & & & & & & & \Checkmark & & & \Checkmark \\
        \hline
        Incremental detection & & & & & & & & & & & & & \Checkmark & & & & \Checkmark & & & \\
        \hline
        Granularity & & & & \Checkmark & & & & & & & & & \Checkmark & & & & & \Checkmark & & \\
        \hline
        Tool & & & & \Checkmark & & & \Checkmark & & & & & & & & & & & \Checkmark & & \\
        \hline
        Open source tool & & & \Checkmark & & & & & & & & & & & & & & & & & \\
        \hline
        Tool dependency & & & & & & & & & & & & & & & & & & \Checkmark & & \\
        \hline
        Clone type & & & \Checkmark & \Checkmark & \Checkmark & & & & & \Checkmark & & & \Checkmark & & & & & \Checkmark & \Checkmark & \Checkmark \\
        \hline
        Program language & & & \Checkmark & \Checkmark & \Checkmark & & & & & & & & & & & & & \Checkmark & & \\
        \hline
        Tool implementation & & & & \Checkmark & & & & & & & & & & & & & & & & \\
        \hline
        IDE & & & & & & & & & & & & & \Checkmark & & & & & & & \\
        \hline
        Tool build metrics & & & & & & & \Checkmark & & & & & & & & & & & & & \\
        \hline
        Evaluation metrics & & & & \Checkmark & & & & & \Checkmark & \Checkmark & & & & & & & & & & \\
        \hline
        Similarity measurement & \Checkmark & & & & \Checkmark & & \Checkmark & & & & & & & & & & & & & \\
        \hline
        Clone mapping & & & & & & \Checkmark & & & & & & & & & & & & & & \\
        \hline
        Evolution pattern & & \Checkmark & & & & & & \Checkmark & & & & & & & & & & & & \Checkmark \\
        \hline
        Genealogy extraction & & & & & & & & & & & & & \Checkmark & & & & & & & \\
        \hline
        GUI support & & & & & & & & & & & & & & & & & & \Checkmark & & \\
        \hline
        Visualization & & & & & & & & & & & & \Checkmark & \Checkmark & & & & & & & \Checkmark \\
        \hline
        Refactor pattern & & & & & & & & & & & & & & & & & & \Checkmark & & \Checkmark \\
        \hline
        \multicolumn{21}{l}{\Checkmark means the paper focused on the research point.}
        \end{tabular}
    \end{center}
\end{table}

For these reasons, the following research questions are posed in this paper:

\textbf{RQ1: What sub-research areas exist for code cloning?}

Code cloning is a popular research area that has received much attention from industrial and academic researchers, but the classification of code cloning sub-research areas is lacking based on scientific research methods and related research topics. We use card sorting to classify and summarize nearly ten years of research on code cloning, obtain sub-research areas and related topics, and make the results of article classification available in the form of an open dataset. The solution of this problem forms a summary of the work in code cloning and facilitates the advancement of subsequent research work.

\textbf{RQ2: What is the development trend of code cloning in general and in each sub-research area?}

The research in code cloning has evolved over a long period, covering various aspects of research, including improvement of detection techniques, maintenance of software quality, and improvement of software development efficiency. Exploring the development trend of code cloning as a whole and each sub-research area can help subsequent researchers to grasp the current status and future development trend of each area and then carry out subsequent research in a targeted manner.

\textbf{RQ3: What is the difference between industry and academic interest in code cloning?}

Code cloning, as a software engineering practice-related issue, has been widely concerned by the industry. Understanding the industry's attention and the change can help subsequent researchers understand the development in the field of practice and clarify the code cloning issues in software development to form a close combination of theory and practice.

\textbf{RQ4: In what form do authors collaborate on research?}

From the researcher's point of view, building a collaborative network exploring the active researchers, research teams, and collaboration patterns can help subsequent researchers follow the relevant research work and build a collaborative relationship.

\textbf{RQ5: Which journals and conferences do code clone papers tend to be published in?}

From the researcher's perspective, the analysis of popular conferences and journals can help subsequent researchers to submit papers in a targeted way. At the same time, they can participate in and follow the work of related research conferences and journals to accelerate the research process.

\section{Systematic literature review}
\begin{figure}[htbp]
  \centering
  \includegraphics[width=\linewidth]{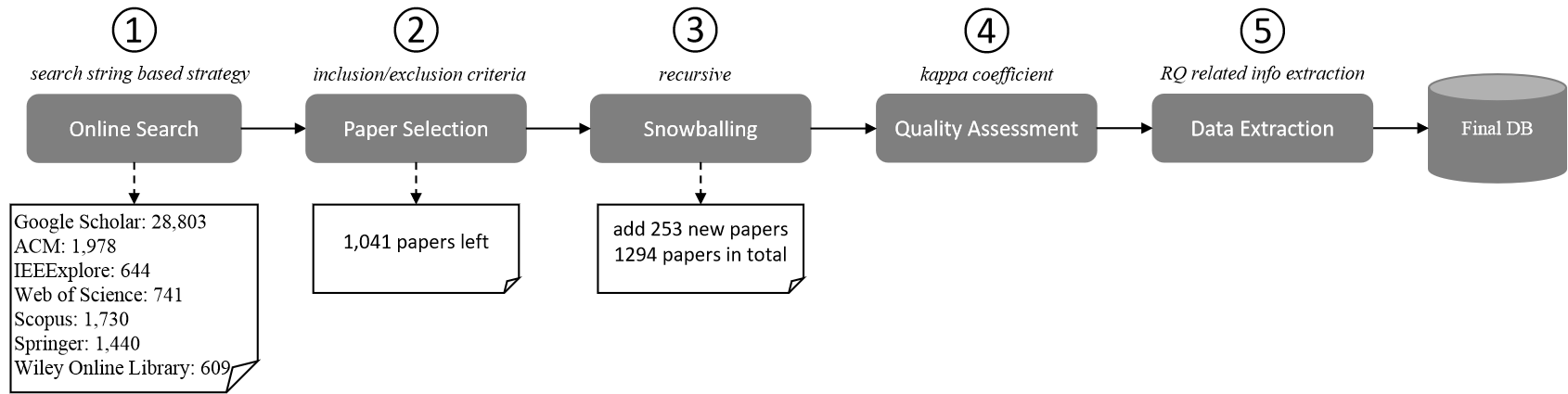}
  \caption{Pipeline of systematic literature review}
  \label{fig1}
\end{figure}

This paper follows the standardized steps of systematic literature review in software engineering~\cite{33}, consisting of five main steps: online search, paper selection, recursive snowballing, quality assessment, and data extraction (Figure~\ref{fig1}).

\subsection{Online search}
\subsubsection{Search method selection}
The current online search methods for systematic literature reviews are divided into two main types: keyword-based search~\cite{33} and target venue-based search~\cite{34}. Both search strategies have their drawbacks. The keyword-based search relies mainly on keywords and search engine selection, where the formation of search terms depends on the research questions and authors' experience~\cite{13,35,36}. This may result in the omission of keyword synonyms, and the selection of search databases and publishers may also omit some relevant articles. For the venue-based search, the selection of targets is crucial, and the absence of target conferences and journals will lead to the loss of a large amount of relevant literature. Although the recursive snowballing method can help reduce the number of missing articles~\cite{37}, it still cannot guarantee the integrity of papers.

Because of the above factors, we finally chose the keyword-based search approach for the following reasons:
\begin{enumerate}
  \item Sobrinho et al.~\cite{34} used the venue-based search method, but in the recursive snowballing phase, their newly discovered 85 articles appeared in 60 uncovered venues;
  \item From WikiCFP~\footnote{http://www.wikicfp.com/cfp/}, we found that there are thousands of journals and conferences under the computer science category, so it is not easy to select target venues to achieve full coverage of papers;
  \item We can optimize our search strategy by combining the keywords defined by the authors in the related work and the selected search database for the keyword-based search method.
\end{enumerate}

\subsubsection{Define search string}
In order to avoid the loss of search keywords, we collected some search keywords from related work, and the results are shown in Table~\ref{table_2}. We can divide the search keywords into two parts:
\begin{enumerate}
  \item Subject: `code', `software', `application'
  \item Action: `clone', `cloning', `copy', `duplicate', `duplication', `similarity', `sibling'
\end{enumerate}
Therefore, we combine the keywords in the subject and action and connect them with OR logic to form the final search string as shown below:

``code clone'' OR ``code cloning'' OR ``code copy'' OR ``code duplicate'' OR ``code duplication'' OR ``code similarity'' OR ``code sibling'' OR ``software clone'' OR ``software cloning'' OR ``software copy'' OR ``software duplicate'' OR ``software duplication'' OR ``software similarity'' OR ``software sibling'' OR ``application clone'' OR ``application cloning'' OR ``application copy'' OR ``application duplicate'' OR ``application duplication'' OR ``application similarity'' OR ``application sibling''

\begin{table}[htbp]
    \begin{center}
        \caption{Search keywords or string in related papers}
        \label{table_2}
        \renewcommand\arraystretch{1.25}
        \footnotesize
        \begin{tabular}{|m{2cm}|m{12cm}|}
        \hline
        \textbf{Paper} & \textbf{Search keywords or strings} \\
        \hline
        \cite{12} & clone; software; code \\
        \hline
        \cite{22} & code clone; code clone tools; code clone prevention; code clone prevention mechanism; code clone management \\
        \hline
        \cite{13} & ((`code' OR `software' OR `application') AND (`clone' OR `cloning' OR `copy' OR `duplicate' OR `duplication' OR `similarity') AND (`change' OR `evolution' OR `genealogy' OR `maintenance' OR `management' OR `tracking')) \\
        \hline
        \cite{38} & code; software; clone; clones; duplication \\
        \hline
        \cite{30} & code clone detection; text-based techniques; tree-based techniques; metric based techniques; PDG based techniques; hybrid techniques; machine learning techniques \\
        \hline
        \cite{25} & ((`code cloning' OR `software code cloning' OR `code cloning tool') AND (`code copy' OR `duplicate code' OR `duplication' OR `code similarity')) \\
        \hline
        \cite{34} & code siblings; copy-and-paste; duplicate code; near-miss clones \\
        \hline
        \end{tabular}
    \end{center}
\end{table}

\subsubsection{Select search database}
By combining the selection of search databases and engines in related works~\cite{12,33,35,39}, the databases searched in this paper include ACM, IEEExplore, Scopus of Elsevier, Springer, Web of Science, Google Scholar, and Wiley Online Library.

\subsubsection{Search tools and methods}
To quickly and accurately retrieve different online resources, we use various tools. For ACM, Springer, and Web of Science, we used the open source Chrome plugin lit-automation/chrome-plugin from GitHub.\footnote{https://github.com/lit-automation/chrome-plugin} For Google Scholar, we used Publish or Perish~\cite{40}. For IEEExplore, Wiley Online Library, and Scopus, we manually searched the query strings in the official websites.

Google Scholar, IEEExplore, and Scopus have a limit on the maximum search string length, so we divide the search string into two separate parts as follows:
\begin{enumerate}
  \item Substring 1: ``code clone'' OR ``code cloning'' OR ``code copy'' OR ``code duplicate'' OR ``code duplication'' OR ``code similarity'' OR ``code sibling'' OR ``software clone'' OR ``software cloning'' OR ``software copy'' OR ``software duplicate'' OR ``software duplication.''
  \item Substring 2: ``software similarity'' OR ``software sibling'' OR ``application clone '' OR ``application cloning'' OR ``application copy'' OR ``application duplicate'' OR ``application duplication'' OR ``application similarity'' OR ``application sibling.''
\end{enumerate}
For the Google Scholar search, the maximum number of results returned is 1000 at a time~\cite{41,42}. In order to cover all relevant articles, we split the search process by year to ensure that each search yields less than 1000 results. (E.g., when searching for Google Scholar results in 2020, the results returned exceeded 1000, so we split substring 1 into 2 strings)

\subsubsection{Search results}
Through the above search strategy, we collected 35,945 papers (search date: 2020-09-13), of which Google Scholar (28,803), ACM (1,978), IEEExplore (644), Web of Science (741), Scopus (1,730), Springer (1,440), Wiley Online Library (609).

\subsection{Paper selection}
\label{paper-selection}
This section sets the inclusion and exclusion rules for paper selection and then filters the papers by manual screening. Here we set up the rules as follows:
\begin{enumerate}
  \item De-duplication. Different search engines and search databases may retrieve the same article. Here we remove duplicate papers according to the title (7,901 articles were removed);
  \item Removal of irrelevant articles. We removed articles that were not related to code cloning by manual checking based on title, abstract, keywords, and full text (24,982 articles were removed);
  \item Removal of non-Chinese or non-English articles (523 articles were removed);
  \item Removal of non-academic research papers. We only retained academic research and removed work such as patents, conference presentations, slides, etc. (1,101 articles were removed);
  \item Removal of work for which the full text could not be found. For subsequent data extraction, we removed work for which the full text could not be found (51 articles were removed);
  \item Paper between 2011 and 2020 was retained. In this paper, we only investigate the development of code cloning in the last decade (346 articles were removed).
\end{enumerate}
The above steps resulted in a total of 1,041 remaining papers.

\subsection{Recursive snowballing}
Since keyword-based search methods have the potential problem of missing synonyms when constructing search strings, to solve this problem, we use the recursive snowballing method to retrieve missing papers in the references of related papers~\cite{37}. Compared with the non-recursive snowballing method~\cite{34}, this method can obtain more relevant articles and thus circumvent the problem of missing papers.
Each stage of recursive snowballing consists of two main steps:
\begin{enumerate}
  \item Automatic reference extraction. Here we use the CERMINE~\cite{44} to extract the reference list of articles automatically;
  \item Determining whether the articles are relevant according to the paper selection rules defined in Section\ref{paper-selection}.
\end{enumerate}
Based on the above steps, no new relevant articles appear after 5 iterations. The paper acquisition and filtering for each iteration are shown in Table~\ref{table_3}.

\begin{table}[htbp]
    \begin{center}
        \caption{The selection of papers for each stage during recursive snowballing}
        \label{table_3}
        \renewcommand\arraystretch{1.25}
        \footnotesize
        \begin{tabular}{m{6cm}|m{1cm}|m{1cm}|m{1cm}|m{1cm}|m{1cm}|m{1cm}}
        \cline{2-7}
         & \multicolumn{6}{c}{Stage} \\
        \cline{2-7}
         & \textbf{1} & \textbf{2} & \textbf{3} & \textbf{4} & \textbf{5} & \textbf{Total} \\
        \hline
        \# extracted papers(-) & 21797 & 4461 & 535 & 41 & 6 & 26840 \\
        \hline
        \# duplications(-) & 15289 & 3100 & 339 & 26 & 5 & 18759 \\
        \hline
        \# unrelated papers(-) & 5217 & 1112 & 166 & 7 & 0 & 6502 \\
        \hline
        \# non-Chinese or non-English papers(-) & 20 & 1 & 0 & 0 & 0 & 21 \\
        \hline
        \# non-research papers(-) & 45 & 10 & 0 & 0 & 0 & 55 \\
        \hline
        \# papers without full text(-) & 74 & 9 & 1 & 0 & 0 & 84 \\
        \hline
        \# papers published before 2011(-) & 928 & 203 & 27 & 7 & 1 & 1166 \\
        \hline
        \# new related papers & 224 & 26 & 2 & 1 & 0 & 253 \\
        \hline
        \multirow{2}{*}{Note} & \multicolumn{6}{l}{Number indicates the number of papers} \\
        \cline{2-7}
         & \multicolumn{6}{l}{(-) indicates the operation of filtering papers} \\
        \hline
        \end{tabular}
    \end{center}
\end{table}

After recursive snowballing, we got 253 new articles. We collected 1294 articles related to code cloning by combining the previous articles.

\subsection{Data extraction}
To assist the subsequent paper topic extraction, sub-research area classification, popularity analysis, industrial-academic research analysis, and author collaboration network analysis, we extracted paper and author information, respectively, and the extracted data included the following:
\begin{enumerate}
  \item Paper information: venue name, publication date, title, abstract, and keywords;
  \item Author information: author's name, affiliation, country, email addresses, and order.
\end{enumerate}

\section{Sub-research area classification method}
\label{section4}
\begin{table}[htbp]
    \begin{center}
        \caption{Classification of sub research areas of code clone in related work}
        \label{table_4}
        \renewcommand\arraystretch{1.25}
        \footnotesize
        \begin{tabular}{|m{2cm}|m{12cm}|}
        \hline
        \textbf{Paper} & \textbf{Sub-research areas} \\
        \hline
        \cite{12} & clone evolution; clone analysis; impact of clones on software quality; clone detection in websites; cloning in related areas; clone detection in aspect oriented programming \\
        \hline
        \cite{18} & clone analysis; clone detection; clone management; clone detection tool evaluation \\
        \hline
        \cite{19} & clone management; clone detection; clone visualization; clone refactoring; clone tracking; linked clone editing; software quality control \\
        \hline
        \cite{20} & clone detection; clone management; clone evolution; clone removal; clone defects; clone visualization; evaluation of clone detection; clone taxonomies; multiple version of clone; plagiarism; copyright infringement; product lines; aspect mining; quality analysis; clone merging; origin analysis; program understanding; other \\
        \hline
        \cite{21} & clone detection; clone analysis; clone maintenance and management; survey and tool evaluation \\
        \hline
        \end{tabular}
    \end{center}
\end{table}

Related work has yielded different classification results for studies in the code cloning sub-research area, as shown in Table~\ref{table_4}. To address the inconsistency in classification, this paper uses a scientific classification method for software engineering, ``card sorting''~\cite{47,48}, based on the synthesis of a more comprehensive collection of code clone related studies from the last decade to classify the code clone sub-research areas. The topics included and the paper correspondence is published as datasets to the GitLink platform.~\footnote{https://www.gitlink.org.cn/Nigel/jos\_code\_clone\_trend\_future/}
In this paper, the classification of code clone sub-research areas based on the card sorting method consists of the following steps:
\begin{enumerate}
  \item Preparation: The first four authors shared the card sorting series by first adding cards to each relevant article based on its title, keywords, abstract information, and even full text. The information on the cards can be summarized or extracted according to the developer's experience or relevant information. The card information includes two parts: the English name of the topic and the Chinese description information (to facilitate the understanding of the topic and speed up the merging of new articles).
  \item Card classification: There are two main methods of card classification. One checks the classification consistency after separate parallel coding; the other is single-threaded common classification~\cite{49}. We adopted the latter method to discuss and agree on disagreements during the classification process~\cite{50}, which can quickly construct domain knowledge, update cognition and form consistent conclusions on time. We used the open coding approach (OCA)~\cite{48}, which generates sub-research domains during the sorting process. The authors integrate topics that focus on the same research area or similar research questions according to the topic description in the cards. E.g., the topic ``license violation detection'' is described as ``Detection of open source license violations based on clone detection''; ``bug localization'' is described as ``locating bugs in code using code cloning techniques''. These topics are applications of clone detection techniques in other fields, so they were merged and replaced by the card ``other fields based on clone detection technique.'' The final integration resulted in all the first-level cards, all the sub-study fields, and all the initial cards, which are the topics included in that sub-study field).
\end{enumerate}

\section{Result}
This section presents the results to answer the five research questions posed in Section~\ref{background}.

\subsection{RQ1: What sub-research areas exist for code cloning?}

\subsubsection{Popularity analysis of sub-research areas}
We obtained the classification results with the card sorting method, as shown in Table~\ref{table_5}. According to Bharti et al.~\cite{51}, clone management covers various fields such as clone retrieval, clone visualization, clone refactoring, clone detection, etc. Therefore, in this paper, only papers related to the proposed clone management framework are integrated under the clone management category while classifying. To facilitate the subsequent research, we set up a separate category of clone survey, which intersects with other categories, i.e., papers related to clone survey belong to other sub-research areas.

From Table~\ref{table_5}, we can see that clone detection has been the most popular sub-research area in the past 10 years. It is the basis for other research areas, including clone analysis, clone evolution, etc. Although clone detection relies on many validation datasets and methods, there are few relevant articles in the two research areas of clone datasets and evaluation. Many clone detection methods rely on the same large-scale datasets and validation algorithms. Based on clone detection methods, many analytical papers and two subfields of clone evolution and clone refactoring have been derived, mainly analyzing and addressing software quality and health. At the same time, we can find that clone detection techniques penetrate many other software-related fields, with 90 research works focusing on other research areas based on clone detection techniques.

\begin{table}[htbp]
    \begin{center}
        \caption{Classification of sub research areas of code clone}
        \label{table_5}
        \renewcommand\arraystretch{1.25}
        \footnotesize
        \begin{tabular}{|m{6cm}|m{1cm}|m{7cm}|}
        \hline
        \textbf{Category} & \textbf{\# papers} & \textbf{Paper description} \\
        \hline
        clone detection technique & 749 & Related papers propose new clone detection methods. \\
        \hline
        clone analysis & 183 & Related papers have conducted empirical studies based on clone detection methods or targeting the field of code cloning (does not include analytical articles dedicated to other sub-research areas). \\
        \hline
        clone evolution & 143 & Related papers focus on the changes in code cloning during the software lifecycle. \\
        \hline
        clone refactoring & 110 & Related papers focus on the refactoring of code clones. \\
        \hline
        other fields based on clone detection technique & 90 & Related papers use clone detection technology to solve problems in other areas. \\
        \hline
        survey and tutorial & 87 & Related papers summarize and analyze relevant articles or technical reports from previous fields. \\
        \hline
        clone visualization & 52 & The main contribution of the related papers is the proposed visualization method or tool for code cloning. \\
        \hline
        clone evaluation & 28 & Related papers describe metrics, tools, methods, etc., for evaluating clone detection techniques. \\
        \hline
        benchmark & 21 & Related papers construct or evaluate datasets on code clone detection. \\
        \hline
        clone management & 14 & Related papers focus on the macro-level of code clone management (e.g., proposing frameworks, mechanisms, concepts, etc.) \\
        \hline
        \end{tabular}
    \end{center}
\end{table}

\subsubsection{Correlation analysis of sub-research areas}
\begin{figure}[htbp]
  \centering
  \includegraphics[width=\linewidth]{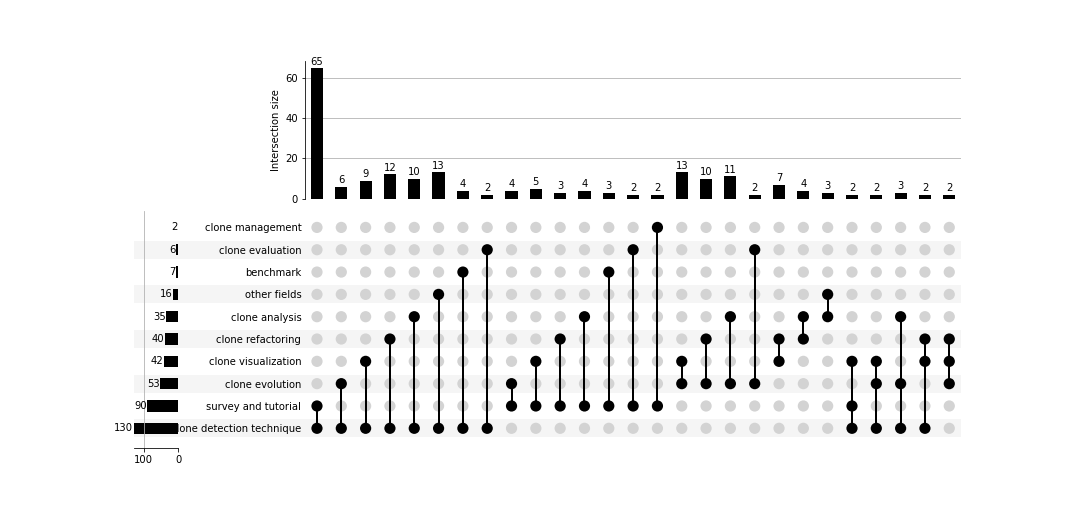}
  \caption{Intersection plot of different sub research fields}
  \label{fig2}
\end{figure}

In our classification, a paper belongs to more than one category, and Figure~\ref{fig2} shows the intersection of papers related to different sub-research areas. For the sake of presentation, only sets containing at least 2 articles are included in the intersection, and the full image has been uploaded to GitLink.~\footnote{https://www.gitlink.org.cn/Nigel/jos\_code\_clone\_trend\_future/tree/master/upsetplot.png} The figure shows that the articles related to code clone research mainly focus on the summary of clone detection techniques (65 out of 87 articles are related to the discussion of clone detection methods). In contrast, for other sub-research areas, except for ``other fields based on cloning detection technique'', all the others have surveys or tutorials. This paper analyzes the relevant surveys in Section~\ref{5.1.3} and summarizes the ``other fields based on cloning detection technique'' in detail. In addition, we found that clone evolution has a strong correlation with clone visualization, clone analysis, and clone refactoring, respectively. More than 10 related papers indicate that visualization support based on clone evolution and software quality-oriented clone analysis and refactoring are the main research directions of clone evolution.

\subsubsection{Analysis of related themes in sub-research areas}
\label{5.1.3}
\begin{figure}[htbp]
  \centering
  \includegraphics[width=\linewidth]{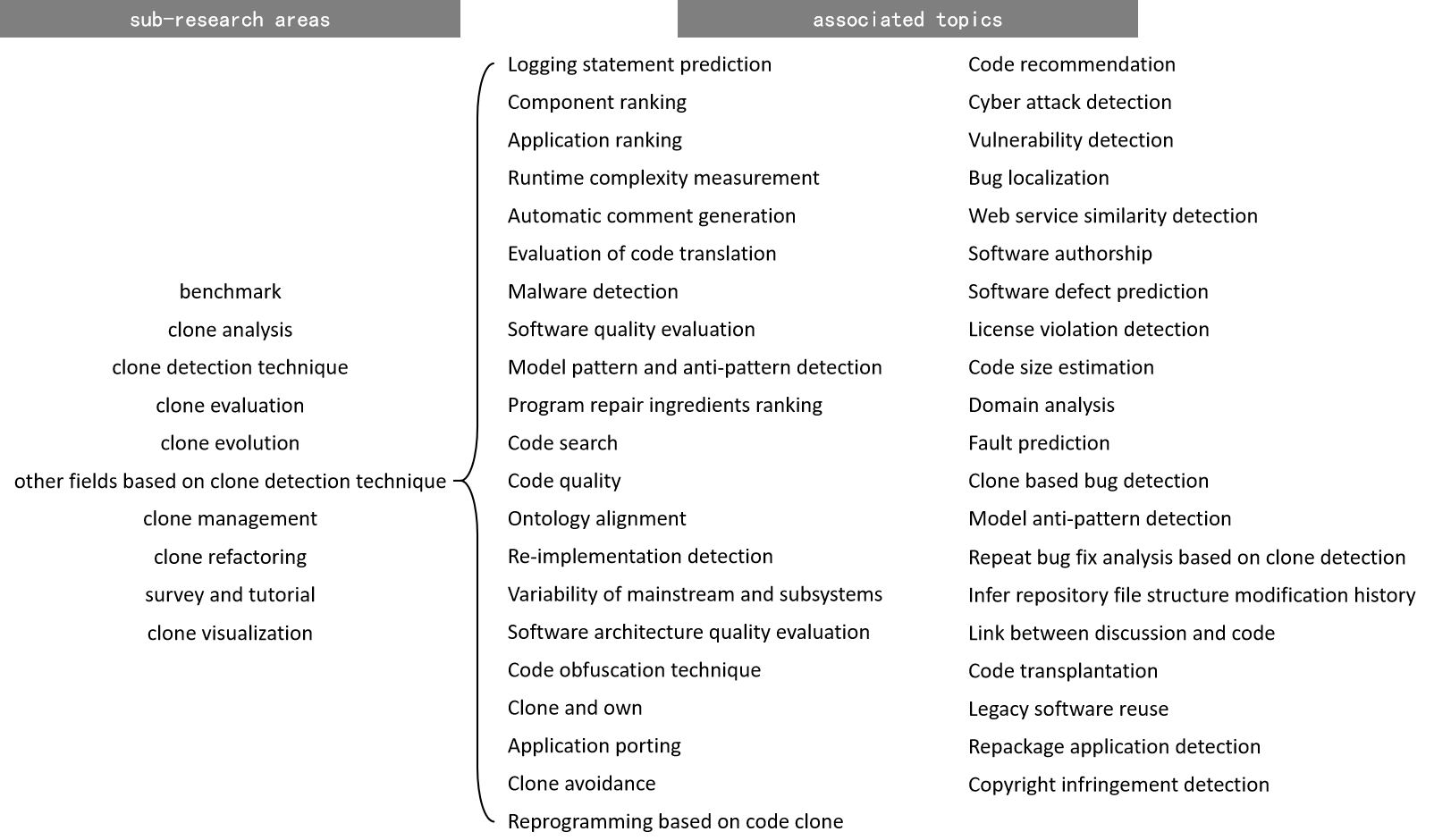}
  \caption{Related topics of other fields based on clone detection technique}
  \label{fig3}
\end{figure}

In card sorting, we set the topics to which the related research belonged. We then integrated the topics to finally form the correspondence between the sub-research areas and their topics (see the description of the card sorting steps in Section~\ref{section4} for the specific process). Figure~\ref{fig3} shows the related research topics of ``other fields based on clone detection technique'', which we present in detail due to the lack of detailed analysis of this area in the related research work. The topic association diagram of all research areas has been uploaded to GitLink in a tree diagram).~\footnote{https://www.gitlink.org.cn/Nigel/jos\_code\_clone\_trend\_future/tree/master/class\_topic\_relation\_tree.html}

We found that all the 90 articles in ``other fields based on clone detection technique'' were classified into 41 topics, including code search (19 related papers), malware detection (13 related papers), and vulnerability detection (10 related papers), which are the three most popular topics. Due to the characteristics of code clone detection, in software engineering, code clone itself can be used as a class of methods to assist in code reuse and software quality management by calculating the similarity between codes. Therefore, code cloning itself has advantages and disadvantages. However, it can increase code maintenance costs~\cite{52}, lead to the propagation of software vulnerabilities~\cite{53}, reduce code readability~\cite{54}, etc. However, at the same time, code clone techniques can also improve the efficiency of software reuse~\cite{55}, speed up software development~\cite{56}, detect software vulnerabilities~\cite{57}, and predict code errors~\cite{58}, etc.

The statistics and analysis of topics contained in sub-research areas can help subsequent researchers quickly build domain knowledge.

\subsection{RQ2: What is the development trend of code cloning in general and in each sub-research area?}
\begin{figure}[htbp]
  \centering
  \includegraphics[width=\linewidth]{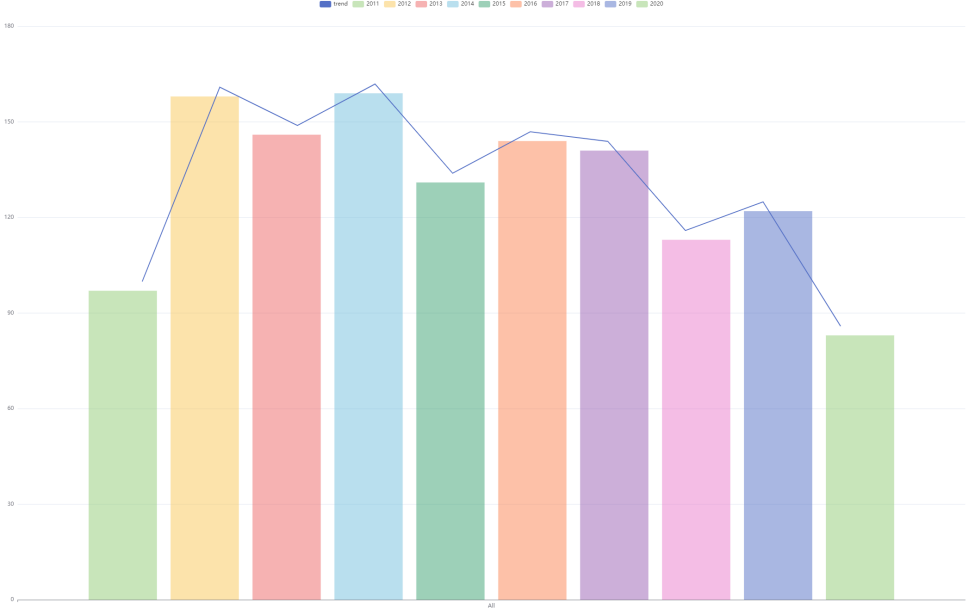}
  \caption{The trend of the overall perspective of code clone}
  \label{fig4}
\end{figure}

\begin{figure}[htbp]
  \centering
  \includegraphics[width=\linewidth]{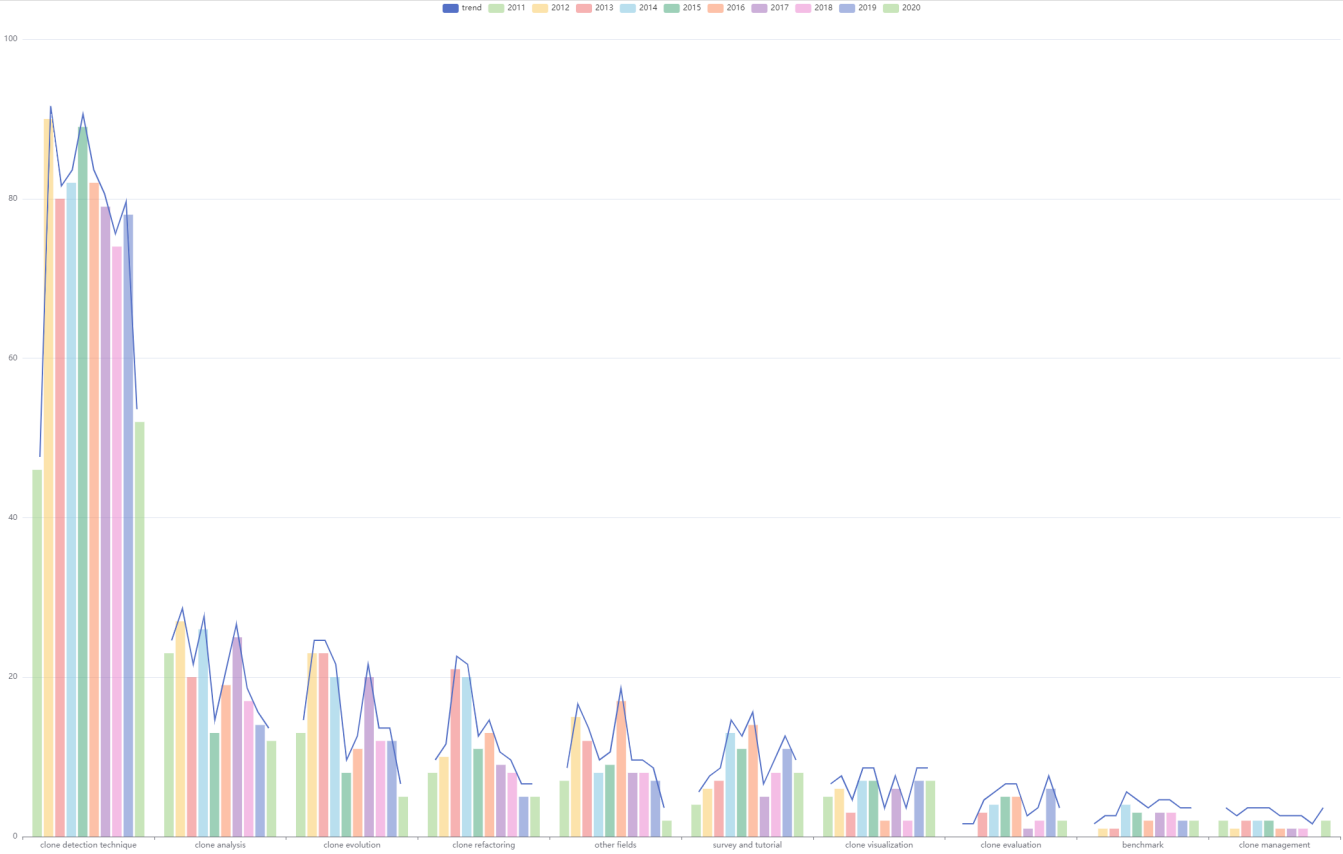}
  \caption{The trend of sub-research fields of code clone}
  \label{fig5}
\end{figure}

We analyzed the development trend of clone detection as a whole and each sub-research area in chronological order. The results are shown in Figure~\ref{fig4} and Figure~\ref{fig5} (see GitLink~\footnote{https://www.gitlink.org.cn/Nigel/jos\_code\_clone\_trend\_future/tree/master/overall\_hotness\_change\_trend.html}\footnote{https://www.gitlink.org.cn/Nigel/jos\_code\_clone\_trend\_future/tree/master/sub-fields\_hotness\_change\_trend.html} for the HTML code of the images). From the overall change of popularity, code cloning experienced a rapid development from 2011 to 2012 and peaked. The subsequent research popularity showed a zigzag development trend. Since the search for related papers was conducted in September 2020, we do not have the complete data for 2020, so we do not consider the popularity of articles related to code cloning in 2020 for the time being. From the sub-research areas, we find that the popularity of clone detection is similar to the overall trend, which is probably because the articles related to clone detection account for a considerable proportion of all articles.

Compared with other sub-research areas, we found that the number of related articles, including clone analysis, clone evolution, and clone refactoring, decreased significantly in 2015 (the opposite trend to clone detection). However, the enthusiasm increased to a certain extent in the next one or two years, probably because the clone detection method is the basis of other sub-research areas. The new clone detection method in 2015 has facilitated the subsequent development of other sub-research areas of code cloning. However, from the overall trend, these three sub-research areas are similar to clone detection, showing an overall decreasing trend.

From the overall trend, the three sub-research areas of clone visualization, clone dataset, and clone management have maintained a stable or rising trend in the past one or two years and have not declined. We think the possible reasons are as follows:
\begin{enumerate}
  \item The dataset is the basis for the formation and breakthrough of the clone detection method, and its development will be a breakthrough due to the bottleneck of the clone detection method. The proposed new dataset means that the clone detection method will have significant progress in some program languages or some aspects;
  \item The proposed clone visualization and clone management framework are clone detection methods in software maintenance and management applications. The development of related work has a certain lag compared with clone detection methods.
\end{enumerate}

\subsection{RQ3: What is the difference between industry and academic interest in code cloning?}
We categorized the papers that included authors affiliated with industry as articles of interest to industry and those that included only academic research institutions as articles of interest to academia.

\begin{figure}[htbp]
  \begin{minipage}[t]{0.45\textwidth}
    \centering
    \includegraphics[width=6cm, height=3.5cm]{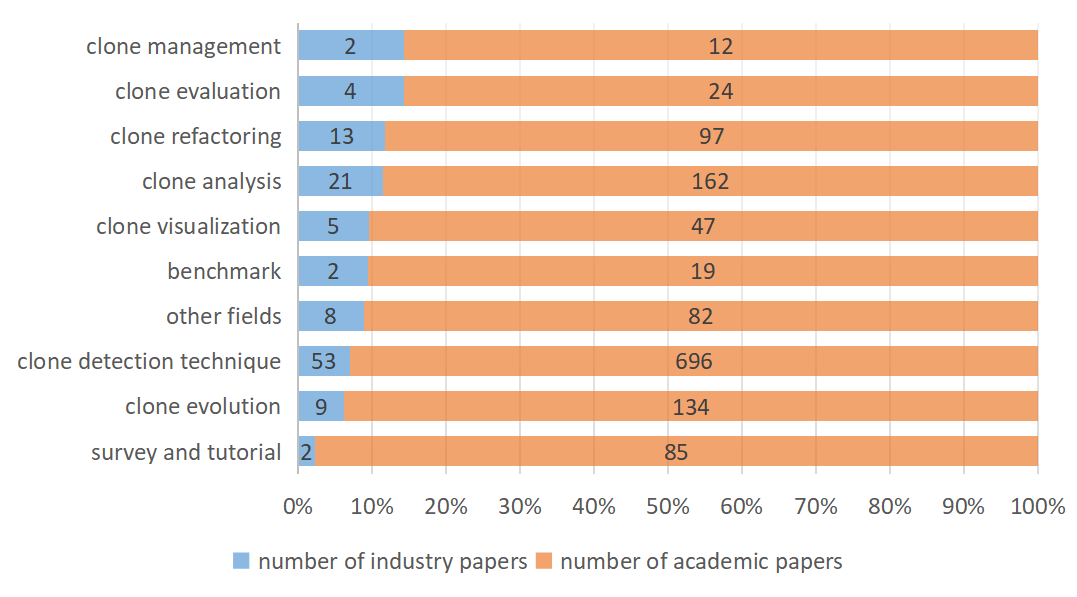}
    \caption{Comparison between industry and academic research hotness of code clone}
    \label{fig6}
  \end{minipage}%
  \hspace{.15in}
  \begin{minipage}[t]{0.45\textwidth}
    \centering
    \includegraphics[width=6cm, height=3.5cm]{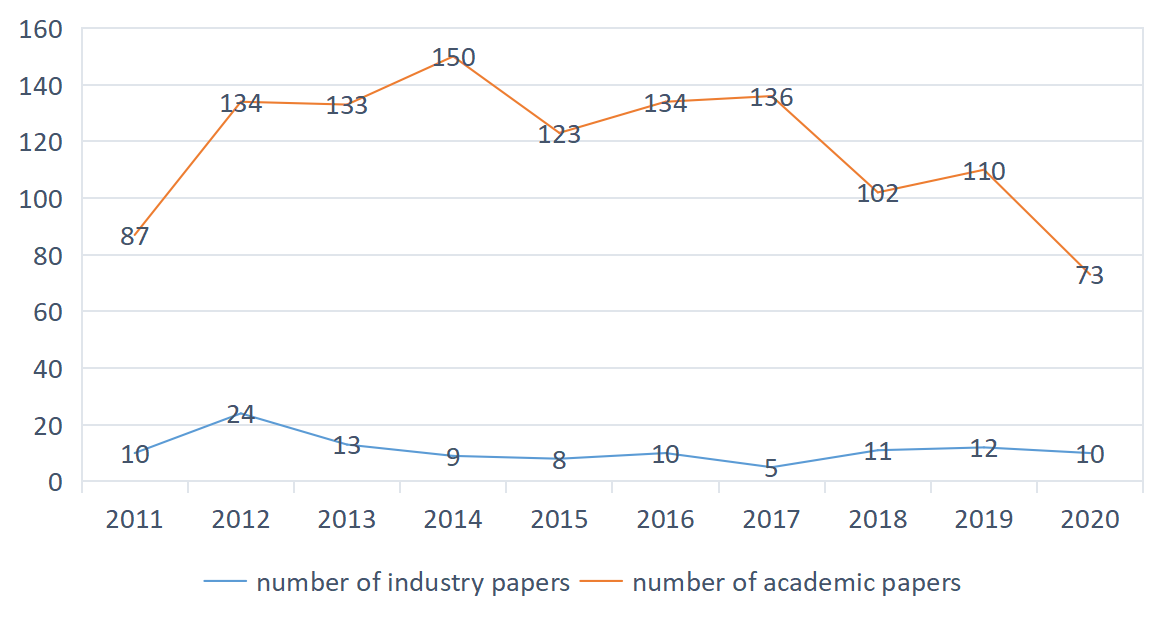}
    \caption{Change of number of industry and academic research papers overtime}
    \label{fig7}
  \end{minipage}
\end{figure}

We found that 112 (8.66\%) of the papers related to code cloning in the past 10 years involved the industry, indicating that the industry has some interest in the field of code cloning. We compared the percentage of relevant articles in each sub-research area (Figure~\ref{fig6}). The figure shows that the industry pays more attention to the sub-research areas of clone management, clone evaluation, clone refactoring, and clone analysis (with a higher percentage of articles). In contrast, the attention to the area of clone detection is relatively low. This shows that the industry is more concerned with the impact of code cloning on software quality and how to refactor and manage code clones.

Comparing the change in the research intensity of code cloning between industry and academia over time (Figure~\ref{fig7}), we find that industry (2012) reached the peak of research intensity sooner than academia (2014). The industry has shown a warming trend for code cloning research in the last 2 to 3 years, while academia has a downward trend in the overall research intensity. The possible reason for this phenomenon is that the problem of code cloning was discovered by industry from a practical problem. Then academia paid attention to the problem and continued to propose new ideas, methods, and results, which were adopted and applied in industry. Therefore, the industry continues to pay attention to the problems related to code cloning.

\subsection{RQ4: In what form do authors collaborate on research?}
\begin{figure}[htbp]
  \centering
  \includegraphics[width=0.8\linewidth]{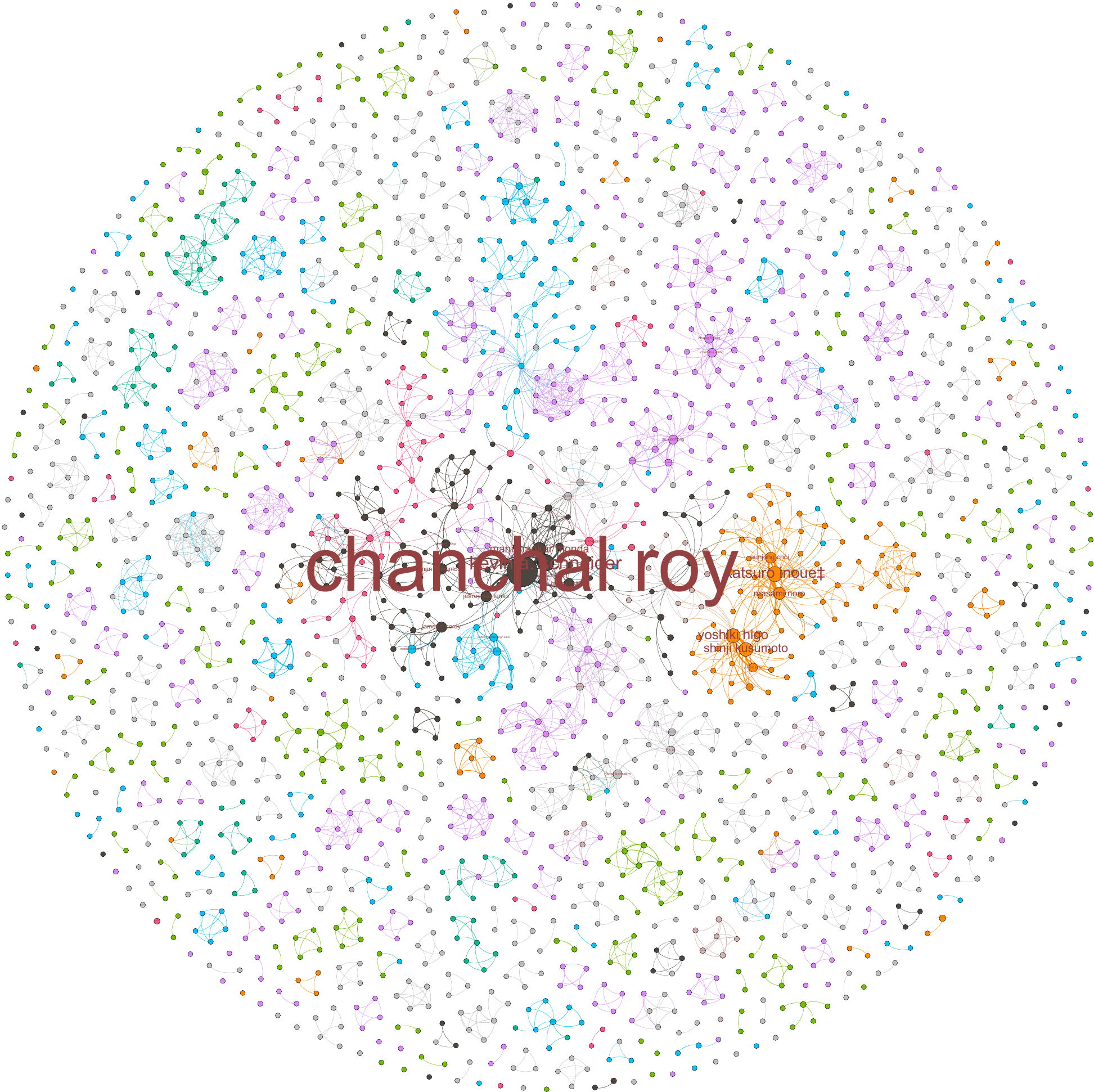}
  \caption{Co-author network of code clone research field}
  \label{fig8}
\end{figure}

Figure~\ref{fig8} shows the collaboration network of authors in code cloning. Nodes indicate authors. Edges indicate that two authors have collaborated in at least one paper. Node size is positively correlated with the number of articles the author has participated in. The size of the author's name corresponding to the node is positively correlated with the node size. The color indicates the country and region to which the author belongs.
The same color indicates the same country or region, and the color depth of the edge is positively correlated with the number of papers co-authored by the related author.

It is found that 72.4\% (1,568) of the authors have only contributed to one article, 4.9\% (106) of the authors have contributed to five or more articles related to code cloning, and the most involved author is Chanchal K. Roy (the largest node as shown in Figure~\ref{fig8}). He has contributed to 95 articles related to code cloning. Thus, although the field of code cloning is a popular research area with a large number of researchers involved in related subfields, only a small number of researchers are focused on this area and continue to contribute research results. We analyzed the characteristics of the authors' cooperative network and found that the average clustering coefficient ($C$) of the whole network was 0.913~\cite{59}, and the average path length ($L$) was 4.906~\cite{60}. Combined with the method proposed by Telesford et al.~\cite{61} for identifying small-world complex networks, we generated an equivalent random-derived network based on the connectivity probability using Gephi~\cite{62}, we obtained the average clustering coefficient ($C_{rand}$) of this random network as 0.001 and the average path length ($L_{rand}$) as 6.332. On this basis, we calculated the small-world coefficients of the authors' cooperative network ($\delta=\frac{C/C_{rand}}{L/L_{rand}}=1178.38$)~\cite{63,64}.
In summary, the cooperative network of code cloning authors satisfies the characteristics of a small-world network, i.e., $C \gg C_{rand},L \approx L_{rand},\delta>1$. From the cooperative network of code cloning authors, it can be seen that the research related to code cloning exhibits the characteristics of a ``small-world network,'' and the authors present the phenomenon of clustering. In addition, the frequency of authors' research is characterized by ``high intensity but low persistence,'' i.e., although a large number of researchers are involved in research related to code cloning, only a few of them are continuously interested in the field.

In terms of the countries the research institutions belonged to, the top countries were China, India, the United States, Canada, and Japan. We counted the node size and graph density of the author partnership network for each country separately (see Table~\ref{table_6}). In terms of the authors' paper output (node size), the teams from the University of Saskatchewan (Chanchal K. Roy, Kevin A. Schneider) in Canada, and Osaka University (Osaka University), Nanzan (representative researchers: Katsuro Inoue, Shinji Kusumoto, Yoshiki Higo, Masami Noro) in Japan, and the joint team of Osaka University and Nanzan University (representative researchers: Katsuro Inoue, Shinji Kusumoto, Yoshiki Higo, Masami Noro) in Japan, have continued to pay attention and made many contributions in the field of code cloning. In comparison, the graph density of the respective subgraphs of China, India, and the United States is relatively low, indicating that the collaboration among authors is not strong. However, the average node degree of our subgraphs is higher than that of India and the United States, indicating that there are some high-producing teams or active authors in China, which increases the overall average collaboration degree.

\begin{table}[htbp]
    \begin{center}
        \caption{Proportion of authors from research institutions of different countries and analysis of co-author networks}
        \label{table_6}
        \renewcommand\arraystretch{1.25}
        \footnotesize
        \begin{tabular}{|m{4cm}|m{1.5cm}|m{1.5cm}|m{2cm}|m{1.5cm}|m{1.5cm}|}
        \hline
         & \textbf{China} & \textbf{India} & \textbf{The United States} & \textbf{Canada} & \textbf{Japan} \\
        \hline
        The percentage of authors(\%) & 24.46 & 15.64 & 10.66 & 6.32 & 6.05 \\
        \hline
        Average node degree & 3.694 & 1.917 & 2.476 & 3.956 & 4.137 \\
        \hline
        Subgraph density & 0.007 & 0.006 & 0.011 & 0.029 & 0.032 \\
        \hline
        \end{tabular}
    \end{center}
\end{table}

To help future researchers better track the popular developers and teams in the relevant sub-research areas, we counted the relevant researchers and the research teams belonging to each sub-research area separately according to the number of published articles (Table~\ref{table_7}).

\begin{table}[htbp]
    \begin{center}
        \caption{Popular researchers and research teams in various sub-research areas}
        \label{table_7}
        \renewcommand\arraystretch{1.25}
        \footnotesize
        \begin{tabular}{|m{4.7cm}|m{8.5cm}|}
        \hline
        \textbf{Sub-research areas} & \textbf{Popular researchers[affiliated teams](\# publications)} \\
        \hline
        \multirow{3}{*}{clone detection} & Chanchal K. Roy[University of Saskatchewan](25); \\
         & Yoshiki Higo[Osaka University](13); \\
         & Oscar Karnalim[University of Newcastle, Maranatha Christian University](12) \\
        \hline
        \multirow{3}{*}{clone evolution} & Chanchal K. Roy[University of Saskatchewan](29); \\
         & Kevin A. Schneider[University of Saskatchewan](25); \\
         & Manishankar Mondal[University of Saskatchewan](22) \\
        \hline
        \multirow{3}{*}{clone refactoring} & Katsuro Inoue[Osaka University](8); \\
         & Chanchal K. Roy[University of Saskatchewan](7); \\
         & Masami Noro[Nanzan University](6) \\
        \hline
        \multirow{3}{*}{other fields based on clone detection technique} & Mohammad Reza Farhadi[Concordia University](5); \\
         & James R. Cordy[Queen's University](5); \\
         & Katsuro Inoue[Osaka University](5) \\
        \hline
        \multirow{3}{*}{survey and tutorial} & Ritu Garg[Indira Gandhi Delhi Technical University for Women](4); \\
         & Chanchal K. Roy[University of Saskatchewan](3); \\
         & Dhavleesh Rattan[Punjabi University](2) \\
        \hline
        \multirow{3}{*}{clone visualization} & Chanchal K. Roy[University of Saskatchewan](7); \\
         & Kevin A. Schneider[University of Saskatchewan](6); \\
         & Katsuro Inoue[Osaka University](5) \\
        \hline
        \multirow{3}{*}{clone evaluation} & Jeffrey Savjlenko[University of Saskatchewan](12); \\
         & Chanchal K. Roy[University of Saskatchewan](11); \\
         & Matthew Stephan[Queen's University, Miami University](4) \\
        \hline
        \multirow{3}{*}{benchmark} & Jeffrey Savjlenko[University of Saskatchewan](3); \\
         & Alan Charpentier[University of Bordeaux](3); \\
         & Chanchal K. Roy[University of Saskatchewan](3) \\
        \hline
        \multirow{3}{*}{clone management} & Minhaz F. Zibran[University of Saskatchewan, University of New Orleans](3); \\
         & Hamid Abdul Basit[Lahore University of Management Sciences](2); \\
         & Jan Harder[University of Bremen](2) \\
        \hline
        \end{tabular}
    \end{center}
\end{table}

From the results, it can be seen that there are differences in the popular researchers and teams in different sub-research areas. However, we found that researchers from the University of Saskatchewan team and Osaka University team actively research several sub-research areas.

\subsection{RQ5: Which journals and conferences do code clone papers tend to be published in?}

\begin{figure}[htbp]
  \centering
  \includegraphics[width=0.8\linewidth]{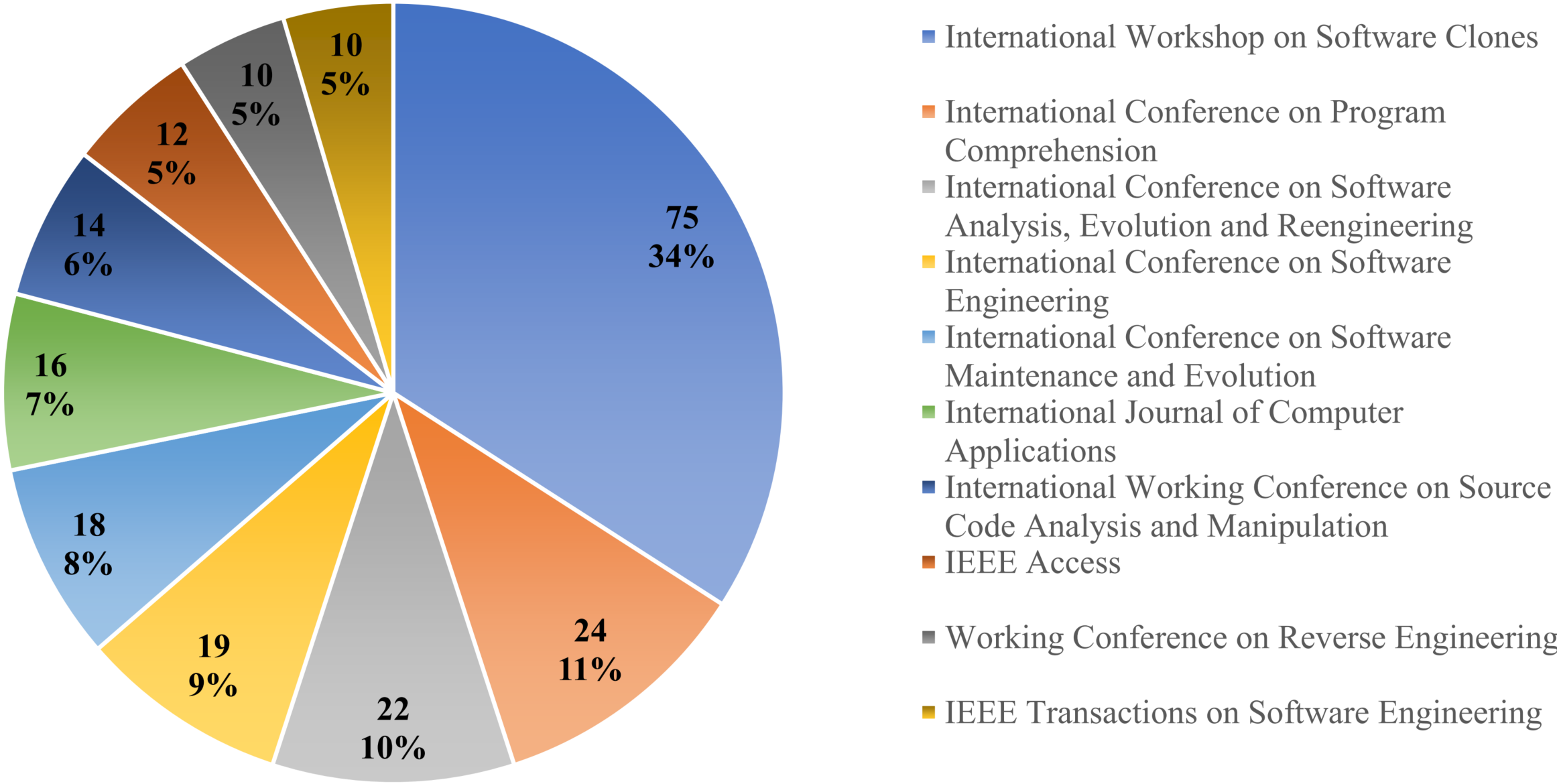}
  \caption{The submission distribution of popular journals and conferences}
  \label{fig9}
\end{figure}

Figure~\ref{fig9} shows the list of the top 10 most popular journals and conferences, including the number of relevant articles published, the percentage of articles, and the names of the corresponding journals and conferences. From the results, we can see that many articles come from the International Workshop on Software Clones, a workshop under the ICSME (International Conference on Software Maintenance and Evolution), which has been held for 15 sessions until 2021. Future researchers may consider this workshop when submitting papers or tracking related conferences.

\section{Discussion}
From the popularity analysis of code detection results, the overall research in this area has a decreasing trend. However, there is an increasing trend for clone visualization, clone management, and related research analysis. At the same time, we found that the industry has gradually increased its attention to code clone-related research since 2017. More attention has been paid to clone management, evaluation, and refactoring than academic sessions. Future researchers can focus on the clone visualization, enrich the relevant tools, improve the relevant management system, and achieve the transformation results.

Code clone detection has been the most popular sub-research area of code cloning as a supporting technology. For this area of research, the optimization directions include accuracy, scalability, and execution efficiency~\cite{65}. Regarding accuracy, we found that existing clone detection methods are very effective for detecting various types of clones~\cite{66}. For this reason, many developers started to analyze the performance of existing clone detection algorithms on particular clone problems (large gap~\cite{67}, large variance clones~\cite{68}) and give corresponding solutions. At the same time, a large amount of related work started to focus on optimizing the execution efficiency of clone detection methods~\cite{69}. Many studies focus on clone detection and analysis of unique codes, e.g., smart contracts~\cite{70}. Future researchers who wish to optimize existing clone detection algorithms can start from execution efficiency and scalability or consider optimizing clone detection techniques in special contexts.

As a key technology, clone detection plays an important role in various fields such as code recommendation, malicious code detection, and software quality assessment. Although there is a decreasing trend of research on ``other fields based on clone detection technique'', we believe that a large amount of research on code analysis, software reuse, etc., can use clone detection as a problem-solving method or optimization technique.

\section{Conclusion}
As an important research area of software engineering, code cloning has received much attention from researchers. Many related studies have been conducted to explore various sub-research areas of code cloning. However, there is a lack of comprehensive introduction and popularity analysis of each sub-research area.
This paper collected 1,294 research papers on code cloning in the past 10 years according to the detailed steps of systematic literature review, and divided the research on code cloning into 10 sub-research areas by card sorting, and did the following analysis:
\begin{enumerate}
  \item We explored the overall popularity of sub-research areas and the intersection of related research areas and uncovered the key research directions and topics of sub-research areas;
  \item We analyzed the change of the overall popularity of code cloning and each sub-research area over time and found the decline of the overall popularity and the difference of the development trend of each sub-research area;
  \item We compared and analyzed the difference of the attention of industry and academia to the sub-research area of code cloning and the change of the attention over time, and found that the industry's tendencies towards clone management and software quality maintenance;
  \item We constructed a network of author collaborations, explored the ``small world'' and ``high intensity but low persistence'' characteristics of code cloning research, discovered the popular researchers and research teams, and analyzed the research institutions in different countries;
  \item We statistically analyzed the popular conferences and journals, which are helpful for the follow-up researchers to submit papers and follow up related research.
\end{enumerate}

This paper discloses the collected papers and classification results, which, together with the results of this paper, can help subsequent researchers quickly build domain knowledge, understand related work, and track research hotspots as well as popular researchers, teams, conferences, and journals. This paper analyzes the research field of code cloning from a macroscopic perspective. Future research work can be guided by the findings and data of this paper to explore each sub-research area more deeply.

\bibliographystyle{ACM-Reference-Format}
\bibliography{sample-base}


\begin{thebibliography}{64}


\ifx \showCODEN    \undefined \def \showCODEN     #1{\unskip}     \fi
\ifx \showDOI      \undefined \def \showDOI       #1{#1}\fi
\ifx \showISBNx    \undefined \def \showISBNx     #1{\unskip}     \fi
\ifx \showISBNxiii \undefined \def \showISBNxiii  #1{\unskip}     \fi
\ifx \showISSN     \undefined \def \showISSN      #1{\unskip}     \fi
\ifx \showLCCN     \undefined \def \showLCCN      #1{\unskip}     \fi
\ifx \shownote     \undefined \def \shownote      #1{#1}          \fi
\ifx \showarticletitle \undefined \def \showarticletitle #1{#1}   \fi
\ifx \showURL      \undefined \def \showURL       {\relax}        \fi
\providecommand\bibfield[2]{#2}
\providecommand\bibinfo[2]{#2}
\providecommand\natexlab[1]{#1}
\providecommand\showeprint[2][]{arXiv:#2}

\bibitem[\protect\citeauthoryear{Ain, Butt, Anwar, Azam, and Maqbool}{Ain
  et~al\mbox{.}}{2019a}]%
        {15}
\bibfield{author}{\bibinfo{person}{Qurat~Ul Ain}, \bibinfo{person}{Wasi~Haider
  Butt}, \bibinfo{person}{Muhamad~Waseem Anwar}, \bibinfo{person}{Farooque
  Azam}, {and} \bibinfo{person}{Bilal Maqbool}.}
  \bibinfo{year}{2019}\natexlab{a}.
\newblock \showarticletitle{Recent advancements in code clone
  detection--techniques and tools}.
\newblock \bibinfo{journal}{\emph{IEEE Access}}  \bibinfo{volume}{7}
  (\bibinfo{year}{2019}).
\newblock


\bibitem[\protect\citeauthoryear{Ain, Butt, Anwar, Azam, and Maqbool}{Ain
  et~al\mbox{.}}{2019b}]%
        {30}
\bibfield{author}{\bibinfo{person}{Qurat~Ul Ain}, \bibinfo{person}{Wasi~Haider
  Butt}, \bibinfo{person}{Muhammad~Waseem Anwar}, \bibinfo{person}{Farooque
  Azam}, {and} \bibinfo{person}{Bilal Maqbool}.}
  \bibinfo{year}{2019}\natexlab{b}.
\newblock \showarticletitle{A systematic review on code clone detection}.
\newblock \bibinfo{journal}{\emph{IEEE access}}  \bibinfo{volume}{7}
  (\bibinfo{year}{2019}), \bibinfo{pages}{86121--86144}.
\newblock


\bibitem[\protect\citeauthoryear{Ali and Sulaiman}{Ali and Sulaiman}{2014}]%
        {22}
\bibfield{author}{\bibinfo{person}{Al-Fahim~Mubarak Ali} {and}
  \bibinfo{person}{Shahida Sulaiman}.} \bibinfo{year}{2014}\natexlab{}.
\newblock \showarticletitle{A systematic literature review of code clone
  prevention approaches}.
\newblock \bibinfo{journal}{\emph{International Journal of Software Engineering
  and Technology}} \bibinfo{volume}{1}, \bibinfo{number}{1}
  (\bibinfo{year}{2014}).
\newblock


\bibitem[\protect\citeauthoryear{Alwaqfi}{Alwaqfi}{2017}]%
        {11}
\bibfield{author}{\bibinfo{person}{Asif Alwaqfi}.}
  \bibinfo{year}{2017}\natexlab{}.
\newblock \emph{\bibinfo{title}{A Refactoring Technique for Large Groups of
  Software Clones}}.
\newblock \bibinfo{thesistype}{Ph.D. Dissertation}. \bibinfo{school}{Concordia
  University}.
\newblock


\bibitem[\protect\citeauthoryear{Auch, Weber, Mandl, and Wolff}{Auch
  et~al\mbox{.}}{2020}]%
        {16}
\bibfield{author}{\bibinfo{person}{Maximilian Auch}, \bibinfo{person}{Manuel
  Weber}, \bibinfo{person}{Peter Mandl}, {and} \bibinfo{person}{Christian
  Wolff}.} \bibinfo{year}{2020}\natexlab{}.
\newblock \showarticletitle{Similarity-based analyses on software applications:
  A systematic literature review}.
\newblock \bibinfo{journal}{\emph{Journal of Systems and Software}}
  \bibinfo{volume}{168} (\bibinfo{year}{2020}), \bibinfo{pages}{110669}.
\newblock


\bibitem[\protect\citeauthoryear{Baars and Oprescu}{Baars and Oprescu}{2019}]%
        {10}
\bibfield{author}{\bibinfo{person}{Simon Baars} {and} \bibinfo{person}{Ana
  Oprescu}.} \bibinfo{year}{2019}\natexlab{}.
\newblock \bibinfo{booktitle}{\emph{Towards automated refactoring of code
  clones in object-oriented programming languages}}.
\newblock \bibinfo{type}{{T}echnical {R}eport}.
  \bibinfo{institution}{EasyChair}.
\newblock


\bibitem[\protect\citeauthoryear{Bandi, Williams, and Allen}{Bandi
  et~al\mbox{.}}{2013}]%
        {35}
\bibfield{author}{\bibinfo{person}{Ajay Bandi}, \bibinfo{person}{Byron~J
  Williams}, {and} \bibinfo{person}{Edward~B Allen}.}
  \bibinfo{year}{2013}\natexlab{}.
\newblock \showarticletitle{Empirical evidence of code decay: A systematic
  mapping study}. In \bibinfo{booktitle}{\emph{2013 20th Working Conference on
  Reverse Engineering (WCRE)}}. IEEE, \bibinfo{pages}{341--350}.
\newblock


\bibitem[\protect\citeauthoryear{Baqais and Alshayeb}{Baqais and
  Alshayeb}{2020}]%
        {39}
\bibfield{author}{\bibinfo{person}{Abdulrahman Ahmed~Bobakr Baqais} {and}
  \bibinfo{person}{Mohammad Alshayeb}.} \bibinfo{year}{2020}\natexlab{}.
\newblock \showarticletitle{Automatic software refactoring: a systematic
  literature review}.
\newblock \bibinfo{journal}{\emph{Software Quality Journal}}
  \bibinfo{volume}{28}, \bibinfo{number}{2} (\bibinfo{year}{2020}),
  \bibinfo{pages}{459--502}.
\newblock


\bibitem[\protect\citeauthoryear{Bastian, Heymann, and Jacomy}{Bastian
  et~al\mbox{.}}{2009}]%
        {62}
\bibfield{author}{\bibinfo{person}{Mathieu Bastian}, \bibinfo{person}{Sebastien
  Heymann}, {and} \bibinfo{person}{Mathieu Jacomy}.}
  \bibinfo{year}{2009}\natexlab{}.
\newblock \showarticletitle{Gephi: an open source software for exploring and
  manipulating networks}. In \bibinfo{booktitle}{\emph{Proceedings of the
  international AAAI conference on web and social media}},
  Vol.~\bibinfo{volume}{3}. \bibinfo{pages}{361--362}.
\newblock


\bibitem[\protect\citeauthoryear{Bazrafshan}{Bazrafshan}{2012}]%
        {6}
\bibfield{author}{\bibinfo{person}{Saman Bazrafshan}.}
  \bibinfo{year}{2012}\natexlab{}.
\newblock \showarticletitle{Evolution of near-miss clones}. In
  \bibinfo{booktitle}{\emph{2012 IEEE 12th International Working Conference on
  Source Code Analysis and Manipulation}}. IEEE, \bibinfo{pages}{74--83}.
\newblock


\bibitem[\protect\citeauthoryear{Begel and Zimmermann}{Begel and
  Zimmermann}{2014}]%
        {49}
\bibfield{author}{\bibinfo{person}{Andrew Begel} {and} \bibinfo{person}{Thomas
  Zimmermann}.} \bibinfo{year}{2014}\natexlab{}.
\newblock \showarticletitle{Analyze this! 145 questions for data scientists in
  software engineering}. In \bibinfo{booktitle}{\emph{Proceedings of the 36th
  International Conference on Software Engineering}}. \bibinfo{pages}{12--23}.
\newblock


\bibitem[\protect\citeauthoryear{Bharti and Singh}{Bharti and Singh}{2020a}]%
        {19}
\bibfield{author}{\bibinfo{person}{Sarveshwar Bharti} {and}
  \bibinfo{person}{Hardeep Singh}.} \bibinfo{year}{2020}\natexlab{a}.
\newblock \showarticletitle{Proactively managing clones inside an IDE: a
  systematic literature review}.
\newblock \bibinfo{journal}{\emph{International Journal of Computers and
  Applications}} (\bibinfo{year}{2020}), \bibinfo{pages}{1--20}.
\newblock


\bibitem[\protect\citeauthoryear{Bharti and Singh}{Bharti and Singh}{2020b}]%
        {51}
\bibfield{author}{\bibinfo{person}{Sarveshwar Bharti} {and}
  \bibinfo{person}{Hardeep Singh}.} \bibinfo{year}{2020}\natexlab{b}.
\newblock \showarticletitle{Proactively managing clones inside an IDE: a
  systematic literature review}.
\newblock \bibinfo{journal}{\emph{International Journal of Computers and
  Applications}} (\bibinfo{year}{2020}), \bibinfo{pages}{1--20}.
\newblock


\bibitem[\protect\citeauthoryear{Bouma}{Bouma}{2012}]%
        {7}
\bibfield{author}{\bibinfo{person}{G Bouma}.} \bibinfo{year}{2012}\natexlab{}.
\newblock \showarticletitle{Studying the Effects of Code Clone Size on Clone
  Evolution}.
\newblock  (\bibinfo{year}{2012}).
\newblock


\bibitem[\protect\citeauthoryear{Brandes}{Brandes}{2001}]%
        {60}
\bibfield{author}{\bibinfo{person}{Ulrik Brandes}.}
  \bibinfo{year}{2001}\natexlab{}.
\newblock \showarticletitle{A faster algorithm for betweenness centrality}.
\newblock \bibinfo{journal}{\emph{Journal of mathematical sociology}}
  \bibinfo{volume}{25}, \bibinfo{number}{2} (\bibinfo{year}{2001}),
  \bibinfo{pages}{163--177}.
\newblock


\bibitem[\protect\citeauthoryear{Campbell, Quincy, Osserman, and
  Pedersen}{Campbell et~al\mbox{.}}{2013}]%
        {50}
\bibfield{author}{\bibinfo{person}{John~L Campbell}, \bibinfo{person}{Charles
  Quincy}, \bibinfo{person}{Jordan Osserman}, {and} \bibinfo{person}{Ove~K
  Pedersen}.} \bibinfo{year}{2013}\natexlab{}.
\newblock \showarticletitle{Coding in-depth semistructured interviews: Problems
  of unitization and intercoder reliability and agreement}.
\newblock \bibinfo{journal}{\emph{Sociological methods \& research}}
  \bibinfo{volume}{42}, \bibinfo{number}{3} (\bibinfo{year}{2013}),
  \bibinfo{pages}{294--320}.
\newblock


\bibitem[\protect\citeauthoryear{Chatterji}{Chatterji}{2014}]%
        {32}
\bibfield{author}{\bibinfo{person}{Debarshi Chatterji}.}
  \bibinfo{year}{2014}\natexlab{}.
\newblock \bibinfo{booktitle}{\emph{Empirical investigation of causes and
  effects of code clones}}.
\newblock \bibinfo{publisher}{The University of Alabama}.
\newblock


\bibitem[\protect\citeauthoryear{Chen, Li, Yan, and Xia}{Chen
  et~al\mbox{.}}{2019}]%
        {1}
\bibfield{author}{\bibinfo{person}{Qiuyuan Chen}, \bibinfo{person}{Shanping
  Li}, \bibinfo{person}{Meng Yan}, {and} \bibinfo{person}{Xin Xia}.}
  \bibinfo{year}{2019}\natexlab{}.
\newblock \showarticletitle{Code Clone Detection: A Literature Review}.
\newblock \bibinfo{journal}{\emph{Journal of software}} \bibinfo{volume}{30},
  \bibinfo{number}{4} (\bibinfo{year}{2019}), \bibinfo{pages}{962--980}.
\newblock


\bibitem[\protect\citeauthoryear{Chochlov}{Chochlov}{2017}]%
        {27}
\bibfield{author}{\bibinfo{person}{Muslim Chochlov}.}
  \bibinfo{year}{2017}\natexlab{}.
\newblock \showarticletitle{State-of-the-Art Report on Clone Detection}.
\newblock  (\bibinfo{year}{2017}).
\newblock


\bibitem[\protect\citeauthoryear{Cordy and Roy}{Cordy and Roy}{2011}]%
        {4}
\bibfield{author}{\bibinfo{person}{James~R Cordy} {and}
  \bibinfo{person}{Chanchal~K Roy}.} \bibinfo{year}{2011}\natexlab{}.
\newblock \showarticletitle{The NiCad clone detector}. In
  \bibinfo{booktitle}{\emph{2011 IEEE 19th International Conference on Program
  Comprehension}}. IEEE, \bibinfo{pages}{219--220}.
\newblock


\bibitem[\protect\citeauthoryear{de~Paulo~Sobrinho, De~Lucia, and
  de~Almeida~Maia}{de~Paulo~Sobrinho et~al\mbox{.}}{2018}]%
        {34}
\bibfield{author}{\bibinfo{person}{Elder~Vicente de Paulo~Sobrinho},
  \bibinfo{person}{Andrea De~Lucia}, {and} \bibinfo{person}{Marcelo de
  Almeida~Maia}.} \bibinfo{year}{2018}\natexlab{}.
\newblock \showarticletitle{A systematic literature review on bad smells--5
  w's: which, when, what, who, where}.
\newblock \bibinfo{journal}{\emph{IEEE Transactions on Software Engineering}}
  \bibinfo{volume}{47}, \bibinfo{number}{1} (\bibinfo{year}{2018}),
  \bibinfo{pages}{17--66}.
\newblock


\bibitem[\protect\citeauthoryear{Gusenbauer}{Gusenbauer}{2019}]%
        {42}
\bibfield{author}{\bibinfo{person}{Michael Gusenbauer}.}
  \bibinfo{year}{2019}\natexlab{}.
\newblock \showarticletitle{Google Scholar to overshadow them all? Comparing
  the sizes of 12 academic search engines and bibliographic databases}.
\newblock \bibinfo{journal}{\emph{Scientometrics}} \bibinfo{volume}{118},
  \bibinfo{number}{1} (\bibinfo{year}{2019}), \bibinfo{pages}{177--214}.
\newblock


\bibitem[\protect\citeauthoryear{Hammad, Basit, Jarzabek, and Koschke}{Hammad
  et~al\mbox{.}}{2020}]%
        {14}
\bibfield{author}{\bibinfo{person}{Muhammad Hammad},
  \bibinfo{person}{Hamid~Abdul Basit}, \bibinfo{person}{Stan Jarzabek}, {and}
  \bibinfo{person}{Rainer Koschke}.} \bibinfo{year}{2020}\natexlab{}.
\newblock \showarticletitle{A systematic mapping study of clone visualization}.
\newblock \bibinfo{journal}{\emph{Computer Science Review}}
  \bibinfo{volume}{37} (\bibinfo{year}{2020}), \bibinfo{pages}{100266}.
\newblock


\bibitem[\protect\citeauthoryear{Harzing}{Harzing}{2010}]%
        {40}
\bibfield{author}{\bibinfo{person}{Anne-Wil Harzing}.}
  \bibinfo{year}{2010}\natexlab{}.
\newblock \bibinfo{booktitle}{\emph{The publish or perish book}}.
\newblock \bibinfo{publisher}{Tarma Software Research Pty Limited Melbourne,
  Australia}.
\newblock


\bibitem[\protect\citeauthoryear{Harzing}{Harzing}{2012}]%
        {41}
\bibfield{author}{\bibinfo{person}{A.~W. Harzing}.}
  \bibinfo{year}{2012}\natexlab{}.
\newblock \showarticletitle{The Publish or Perish Book: Your Guide to Effective
  and Responsible Citation Analysis}.
\newblock \bibinfo{journal}{\emph{International Review of Research in Open \&
  Distance Learning}} \bibinfo{volume}{13}, \bibinfo{number}{3}
  (\bibinfo{year}{2012}), \bibinfo{pages}{314--315}.
\newblock


\bibitem[\protect\citeauthoryear{Honda, Tokui, Yokoi, Choi, Yoshida, and
  Inoue}{Honda et~al\mbox{.}}{2019}]%
        {8}
\bibfield{author}{\bibinfo{person}{Hirotaka Honda}, \bibinfo{person}{Shogo
  Tokui}, \bibinfo{person}{Kazuki Yokoi}, \bibinfo{person}{Eunjong Choi},
  \bibinfo{person}{Norihiro Yoshida}, {and} \bibinfo{person}{Katsuro Inoue}.}
  \bibinfo{year}{2019}\natexlab{}.
\newblock \showarticletitle{CCEvovis: A clone evolution visualization system
  for software maintenance}. In \bibinfo{booktitle}{\emph{2019 IEEE/ACM 27th
  International Conference on Program Comprehension (ICPC)}}. IEEE,
  \bibinfo{pages}{122--125}.
\newblock


\bibitem[\protect\citeauthoryear{Hordijk, Ponisio, and Wieringa}{Hordijk
  et~al\mbox{.}}{2009}]%
        {38}
\bibfield{author}{\bibinfo{person}{Wiebe Hordijk},
  \bibinfo{person}{Mar{\'\i}a~Laura Ponisio}, {and} \bibinfo{person}{Roel
  Wieringa}.} \bibinfo{year}{2009}\natexlab{}.
\newblock \showarticletitle{Review of code clone articles}.
\newblock \bibinfo{journal}{\emph{University of Twente, The Netherlands,
  http://eprints. eemcs. utwente. nl/12257/01/TR-CTIT-08-33. pdf, accessed on
  May}}  \bibinfo{volume}{18} (\bibinfo{year}{2009}), \bibinfo{pages}{23}.
\newblock


\bibitem[\protect\citeauthoryear{Humphries and Gurney}{Humphries and
  Gurney}{2008}]%
        {64}
\bibfield{author}{\bibinfo{person}{Mark~D Humphries} {and}
  \bibinfo{person}{Kevin Gurney}.} \bibinfo{year}{2008}\natexlab{}.
\newblock \showarticletitle{Network 'small-world-ness': a quantitative method
  for determining canonical network equivalence}.
\newblock \bibinfo{journal}{\emph{PloS one}} \bibinfo{volume}{3},
  \bibinfo{number}{4} (\bibinfo{year}{2008}), \bibinfo{pages}{e0002051}.
\newblock


\bibitem[\protect\citeauthoryear{Humphries, Gurney, and Prescott}{Humphries
  et~al\mbox{.}}{2006}]%
        {63}
\bibfield{author}{\bibinfo{person}{Mark~D Humphries}, \bibinfo{person}{Kevin
  Gurney}, {and} \bibinfo{person}{Tony~J Prescott}.}
  \bibinfo{year}{2006}\natexlab{}.
\newblock \showarticletitle{The brainstem reticular formation is a small-world,
  not scale-free, network}.
\newblock \bibinfo{journal}{\emph{Proceedings of the Royal Society B:
  Biological Sciences}} \bibinfo{volume}{273}, \bibinfo{number}{1585}
  (\bibinfo{year}{2006}), \bibinfo{pages}{503--511}.
\newblock


\bibitem[\protect\citeauthoryear{Jalali and Wohlin}{Jalali and Wohlin}{2012}]%
        {37}
\bibfield{author}{\bibinfo{person}{Samireh Jalali} {and} \bibinfo{person}{Claes
  Wohlin}.} \bibinfo{year}{2012}\natexlab{}.
\newblock \showarticletitle{Systematic literature studies: database searches
  vs. backward snowballing}. In \bibinfo{booktitle}{\emph{Proceedings of the
  2012 ACM-IEEE international symposium on empirical software engineering and
  measurement}}. IEEE, \bibinfo{pages}{29--38}.
\newblock


\bibitem[\protect\citeauthoryear{Kamei, Sato, Monden, Kawaguchi, Uwano, Nagura,
  Matsumoto, and Ubayashi}{Kamei et~al\mbox{.}}{2011}]%
        {58}
\bibfield{author}{\bibinfo{person}{Yasutaka Kamei}, \bibinfo{person}{Hiroki
  Sato}, \bibinfo{person}{Akito Monden}, \bibinfo{person}{Shinji Kawaguchi},
  \bibinfo{person}{Hidetake Uwano}, \bibinfo{person}{Masataka Nagura},
  \bibinfo{person}{Ken-ichi Matsumoto}, {and} \bibinfo{person}{Naoyasu
  Ubayashi}.} \bibinfo{year}{2011}\natexlab{}.
\newblock \showarticletitle{An empirical study of fault prediction with code
  clone metrics}. In \bibinfo{booktitle}{\emph{2011 Joint Conference of the
  21st International Workshop on Software Measurement and the 6th International
  Conference on Software Process and Product Measurement}}. IEEE,
  \bibinfo{pages}{55--61}.
\newblock


\bibitem[\protect\citeauthoryear{Keele et~al\mbox{.}}{Keele
  et~al\mbox{.}}{2007}]%
        {33}
\bibfield{author}{\bibinfo{person}{Staffs Keele} {et~al\mbox{.}}}
  \bibinfo{year}{2007}\natexlab{}.
\newblock \bibinfo{booktitle}{\emph{Guidelines for performing systematic
  literature reviews in software engineering}}.
\newblock \bibinfo{type}{{T}echnical {R}eport}. \bibinfo{institution}{Technical
  report, Ver. 2.3 EBSE Technical Report. EBSE}.
\newblock


\bibitem[\protect\citeauthoryear{Keivanloo, Forbes, and Rilling}{Keivanloo
  et~al\mbox{.}}{2012}]%
        {55}
\bibfield{author}{\bibinfo{person}{Iman Keivanloo},
  \bibinfo{person}{Christopher Forbes}, {and} \bibinfo{person}{Juergen
  Rilling}.} \bibinfo{year}{2012}\natexlab{}.
\newblock \showarticletitle{Similarity search plug-in: Clone detection meets
  internet-scale code search}. In \bibinfo{booktitle}{\emph{2012 4th
  International Workshop on Search-Driven Development: Users, Infrastructure,
  Tools, and Evaluation (SUITE)}}. IEEE, \bibinfo{pages}{21--22}.
\newblock


\bibitem[\protect\citeauthoryear{Koschke}{Koschke}{2008}]%
        {54}
\bibfield{author}{\bibinfo{person}{Rainer Koschke}.}
  \bibinfo{year}{2008}\natexlab{}.
\newblock \showarticletitle{Frontiers of software clone management}. In
  \bibinfo{booktitle}{\emph{2008 Frontiers of Software Maintenance}}. IEEE,
  \bibinfo{pages}{119--128}.
\newblock


\bibitem[\protect\citeauthoryear{Latapy}{Latapy}{2008}]%
        {59}
\bibfield{author}{\bibinfo{person}{Matthieu Latapy}.}
  \bibinfo{year}{2008}\natexlab{}.
\newblock \showarticletitle{Main-memory triangle computations for very large
  (sparse (power-law)) graphs}.
\newblock \bibinfo{journal}{\emph{Theoretical computer science}}
  \bibinfo{volume}{407}, \bibinfo{number}{1-3} (\bibinfo{year}{2008}),
  \bibinfo{pages}{458--473}.
\newblock


\bibitem[\protect\citeauthoryear{Li, Wu, Roy, Sun, Peng, Zhan, Hu, and Ma}{Li
  et~al\mbox{.}}{2020}]%
        {69}
\bibfield{author}{\bibinfo{person}{Guanhua Li}, \bibinfo{person}{Yijian Wu},
  \bibinfo{person}{Chanchal~K Roy}, \bibinfo{person}{Jun Sun},
  \bibinfo{person}{Xin Peng}, \bibinfo{person}{Nanjie Zhan},
  \bibinfo{person}{Bin Hu}, {and} \bibinfo{person}{Jingyi Ma}.}
  \bibinfo{year}{2020}\natexlab{}.
\newblock \showarticletitle{SAGA: efficient and large-scale detection of
  near-miss clones with GPU acceleration}. In \bibinfo{booktitle}{\emph{2020
  IEEE 27th International Conference on Software Analysis, Evolution and
  Reengineering (SANER)}}. IEEE, \bibinfo{pages}{272--283}.
\newblock


\bibitem[\protect\citeauthoryear{Li, Kwon, Kwon, and Lee}{Li
  et~al\mbox{.}}{2016}]%
        {57}
\bibfield{author}{\bibinfo{person}{Hongzhe Li}, \bibinfo{person}{Hyuckmin
  Kwon}, \bibinfo{person}{Jonghoon Kwon}, {and} \bibinfo{person}{Heejo Lee}.}
  \bibinfo{year}{2016}\natexlab{}.
\newblock \showarticletitle{CLORIFI: software vulnerability discovery using
  code clone verification}.
\newblock \bibinfo{journal}{\emph{Concurrency and Computation: Practice and
  Experience}} \bibinfo{volume}{28}, \bibinfo{number}{6}
  (\bibinfo{year}{2016}), \bibinfo{pages}{1900--1917}.
\newblock


\bibitem[\protect\citeauthoryear{Liu, Yang, Jiang, Zhao, and Sun}{Liu
  et~al\mbox{.}}{2019}]%
        {70}
\bibfield{author}{\bibinfo{person}{Han Liu}, \bibinfo{person}{Zhiqiang Yang},
  \bibinfo{person}{Yu Jiang}, \bibinfo{person}{Wenqi Zhao}, {and}
  \bibinfo{person}{Jiaguang Sun}.} \bibinfo{year}{2019}\natexlab{}.
\newblock \showarticletitle{Enabling clone detection for ethereum via smart
  contract birthmarks}. In \bibinfo{booktitle}{\emph{2019 IEEE/ACM 27th
  International Conference on Program Comprehension (ICPC)}}. IEEE,
  \bibinfo{pages}{105--115}.
\newblock


\bibitem[\protect\citeauthoryear{Lozano, Wermelinger, and Nuseibeh}{Lozano
  et~al\mbox{.}}{2007}]%
        {53}
\bibfield{author}{\bibinfo{person}{Angela Lozano}, \bibinfo{person}{Michel
  Wermelinger}, {and} \bibinfo{person}{Bashar Nuseibeh}.}
  \bibinfo{year}{2007}\natexlab{}.
\newblock \showarticletitle{Evaluating the harmfulness of cloning: A change
  based experiment}. In \bibinfo{booktitle}{\emph{Fourth International Workshop
  on Mining Software Repositories (MSR'07: ICSE Workshops 2007)}}. IEEE,
  \bibinfo{pages}{18--18}.
\newblock


\bibitem[\protect\citeauthoryear{Mondal, Roy, and Schneider}{Mondal
  et~al\mbox{.}}{2020}]%
        {31}
\bibfield{author}{\bibinfo{person}{Manishankar Mondal},
  \bibinfo{person}{Chanchal~K Roy}, {and} \bibinfo{person}{Kevin~A Schneider}.}
  \bibinfo{year}{2020}\natexlab{}.
\newblock \showarticletitle{A survey on clone refactoring and tracking}.
\newblock \bibinfo{journal}{\emph{Journal of Systems and Software}}
  \bibinfo{volume}{159} (\bibinfo{year}{2020}), \bibinfo{pages}{110429}.
\newblock


\bibitem[\protect\citeauthoryear{Monden, Nakae, Kamiya, Sato, and
  Matsumoto}{Monden et~al\mbox{.}}{2002}]%
        {52}
\bibfield{author}{\bibinfo{person}{Akito Monden}, \bibinfo{person}{Daikai
  Nakae}, \bibinfo{person}{Toshihiro Kamiya}, \bibinfo{person}{Shin-ichi Sato},
  {and} \bibinfo{person}{Ken-ichi Matsumoto}.} \bibinfo{year}{2002}\natexlab{}.
\newblock \showarticletitle{Software quality analysis by code clones in
  industrial legacy software}. In \bibinfo{booktitle}{\emph{Proceedings Eighth
  IEEE Symposium on Software Metrics}}. IEEE, \bibinfo{pages}{87--94}.
\newblock


\bibitem[\protect\citeauthoryear{Murakami, Higo, and Kusumoto}{Murakami
  et~al\mbox{.}}{2015}]%
        {9}
\bibfield{author}{\bibinfo{person}{Hiroaki Murakami}, \bibinfo{person}{Yoshiki
  Higo}, {and} \bibinfo{person}{Shinji Kusumoto}.}
  \bibinfo{year}{2015}\natexlab{}.
\newblock \showarticletitle{ClonePacker: A tool for clone set visualization}.
  In \bibinfo{booktitle}{\emph{2015 IEEE 22nd International Conference on
  Software Analysis, Evolution, and Reengineering (SANER)}}. IEEE,
  \bibinfo{pages}{474--478}.
\newblock


\bibitem[\protect\citeauthoryear{Nakagawa, Higo, and Kusumoto}{Nakagawa
  et~al\mbox{.}}{2021}]%
        {68}
\bibfield{author}{\bibinfo{person}{Tasuku Nakagawa}, \bibinfo{person}{Yoshiki
  Higo}, {and} \bibinfo{person}{Shinji Kusumoto}.}
  \bibinfo{year}{2021}\natexlab{}.
\newblock \showarticletitle{NIL: large-scale detection of large-variance
  clones}. In \bibinfo{booktitle}{\emph{Proceedings of the 29th ACM Joint
  Meeting on European Software Engineering Conference and Symposium on the
  Foundations of Software Engineering}}. \bibinfo{pages}{830--841}.
\newblock


\bibitem[\protect\citeauthoryear{Paiva and Figueiredo}{Paiva and
  Figueiredo}{[n.d.]}]%
        {28}
\bibfield{author}{\bibinfo{person}{A. Paiva} {and} \bibinfo{person}{E.
  Figueiredo}.} \bibinfo{year}{[n.d.]}\natexlab{}.
\newblock \showarticletitle{Do Concern Metrics Support Code Clone Detection?}
\newblock  (\bibinfo{year}{[n.\,d.]}).
\newblock


\bibitem[\protect\citeauthoryear{Pate, Tairas, and Kraft}{Pate
  et~al\mbox{.}}{2013}]%
        {13}
\bibfield{author}{\bibinfo{person}{Jeremy~R Pate}, \bibinfo{person}{Robert
  Tairas}, {and} \bibinfo{person}{Nicholas~A Kraft}.}
  \bibinfo{year}{2013}\natexlab{}.
\newblock \showarticletitle{Clone evolution: a systematic review}.
\newblock \bibinfo{journal}{\emph{Journal of software: Evolution and Process}}
  \bibinfo{volume}{25}, \bibinfo{number}{3} (\bibinfo{year}{2013}),
  \bibinfo{pages}{261--283}.
\newblock


\bibitem[\protect\citeauthoryear{Patil, Patil, Shinde, and Joshi}{Patil
  et~al\mbox{.}}{2014}]%
        {25}
\bibfield{author}{\bibinfo{person}{Ritesh~V Patil}, \bibinfo{person}{Lalit~V
  Patil}, \bibinfo{person}{Sachin~V Shinde}, {and} \bibinfo{person}{SD Joshi}.}
  \bibinfo{year}{2014}\natexlab{}.
\newblock \showarticletitle{Software code cloning detection and future scope
  development-Latest short review}. In \bibinfo{booktitle}{\emph{International
  Conference on Recent Advances and Innovations in Engineering (ICRAIE-2014)}}.
  IEEE, \bibinfo{pages}{1--4}.
\newblock


\bibitem[\protect\citeauthoryear{Rattan, Bhatia, and Singh}{Rattan
  et~al\mbox{.}}{2013}]%
        {12}
\bibfield{author}{\bibinfo{person}{Dhavleesh Rattan}, \bibinfo{person}{Rajesh
  Bhatia}, {and} \bibinfo{person}{Maninder Singh}.}
  \bibinfo{year}{2013}\natexlab{}.
\newblock \showarticletitle{Software clone detection: A systematic review}.
\newblock \bibinfo{journal}{\emph{Information and Software Technology}}
  \bibinfo{volume}{55}, \bibinfo{number}{7} (\bibinfo{year}{2013}),
  \bibinfo{pages}{1165--1199}.
\newblock


\bibitem[\protect\citeauthoryear{Rattan and Kaur}{Rattan and Kaur}{2016}]%
        {24}
\bibfield{author}{\bibinfo{person}{Dhavleesh Rattan} {and}
  \bibinfo{person}{Jagdeep Kaur}.} \bibinfo{year}{2016}\natexlab{}.
\newblock \showarticletitle{Systematic mapping study of metrics based clone
  detection techniques}. In \bibinfo{booktitle}{\emph{Proceedings of the
  International Conference on Advances in Information Communication Technology
  \& Computing}}. \bibinfo{pages}{1--7}.
\newblock


\bibitem[\protect\citeauthoryear{Sajnani}{Sajnani}{2016}]%
        {65}
\bibfield{author}{\bibinfo{person}{H. Sajnani}.}
  \bibinfo{year}{2016}\natexlab{}.
\newblock \emph{\bibinfo{title}{Large-Scale Code Clone Detection.}}
\newblock \bibinfo{thesistype}{Ph.D. Dissertation}. \bibinfo{school}{University
  of California, Irvine.}
\newblock


\bibitem[\protect\citeauthoryear{Sajnani, Saini, Svajlenko, Roy, and
  Lopes}{Sajnani et~al\mbox{.}}{2016}]%
        {5}
\bibfield{author}{\bibinfo{person}{Hitesh Sajnani}, \bibinfo{person}{Vaibhav
  Saini}, \bibinfo{person}{Jeffrey Svajlenko}, \bibinfo{person}{Chanchal~K
  Roy}, {and} \bibinfo{person}{Cristina~V Lopes}.}
  \bibinfo{year}{2016}\natexlab{}.
\newblock \showarticletitle{Sourcerercc: Scaling code clone detection to
  big-code}. In \bibinfo{booktitle}{\emph{Proceedings of the 38th International
  Conference on Software Engineering}}. \bibinfo{pages}{1157--1168}.
\newblock


\bibitem[\protect\citeauthoryear{Shippey, Bowes, Chrisianson, and Hall}{Shippey
  et~al\mbox{.}}{2012}]%
        {20}
\bibfield{author}{\bibinfo{person}{Thomas Shippey}, \bibinfo{person}{David
  Bowes}, \bibinfo{person}{Bruce Chrisianson}, {and} \bibinfo{person}{Tracy
  Hall}.} \bibinfo{year}{2012}\natexlab{}.
\newblock \showarticletitle{A mapping study of software code cloning}. In
  \bibinfo{booktitle}{\emph{16th International Conference on Evaluation \&
  Assessment in Software Engineering (EASE 2012)}}. IET,
  \bibinfo{pages}{274--278}.
\newblock


\bibitem[\protect\citeauthoryear{Spencer}{Spencer}{2009}]%
        {47}
\bibfield{author}{\bibinfo{person}{Donna Spencer}.}
  \bibinfo{year}{2009}\natexlab{}.
\newblock \bibinfo{booktitle}{\emph{Card sorting: Designing usable
  categories}}.
\newblock \bibinfo{publisher}{Rosenfeld Media}.
\newblock


\bibitem[\protect\citeauthoryear{Su and Zhang}{Su and Zhang}{2018}]%
        {21}
\bibfield{author}{\bibinfo{person}{Xiaohong Su} {and} \bibinfo{person}{Fanlong
  Zhang}.} \bibinfo{year}{2018}\natexlab{}.
\newblock \showarticletitle{A Survey for Management-Oriented Code Clone
  Research}.
\newblock \bibinfo{journal}{\emph{Chinese Journal of Computers}}
  \bibinfo{volume}{41}, \bibinfo{number}{3} (\bibinfo{year}{2018}),
  \bibinfo{pages}{24}.
\newblock


\bibitem[\protect\citeauthoryear{Svajlenko and Roy}{Svajlenko and Roy}{2020}]%
        {26}
\bibfield{author}{\bibinfo{person}{Jeffrey Svajlenko} {and}
  \bibinfo{person}{Chanchal~K Roy}.} \bibinfo{year}{2020}\natexlab{}.
\newblock \showarticletitle{A Survey on the Evaluation of Clone Detection
  Performance and Benchmarking}.
\newblock \bibinfo{journal}{\emph{arXiv preprint arXiv:2006.15682}}
  (\bibinfo{year}{2020}).
\newblock


\bibitem[\protect\citeauthoryear{Telesford, Joyce, Hayasaka, Burdette, and
  Laurienti}{Telesford et~al\mbox{.}}{2011}]%
        {61}
\bibfield{author}{\bibinfo{person}{Qawi~K Telesford}, \bibinfo{person}{Karen~E
  Joyce}, \bibinfo{person}{Satoru Hayasaka}, \bibinfo{person}{Jonathan~H
  Burdette}, {and} \bibinfo{person}{Paul~J Laurienti}.}
  \bibinfo{year}{2011}\natexlab{}.
\newblock \showarticletitle{The ubiquity of small-world networks}.
\newblock \bibinfo{journal}{\emph{Brain connectivity}} \bibinfo{volume}{1},
  \bibinfo{number}{5} (\bibinfo{year}{2011}), \bibinfo{pages}{367--375}.
\newblock


\bibitem[\protect\citeauthoryear{Tkaczyk, Szostek, Fedoryszak, Dendek, and
  Bolikowski}{Tkaczyk et~al\mbox{.}}{2015}]%
        {44}
\bibfield{author}{\bibinfo{person}{Dominika Tkaczyk}, \bibinfo{person}{Pawe{\l}
  Szostek}, \bibinfo{person}{Mateusz Fedoryszak}, \bibinfo{person}{Piotr~Jan
  Dendek}, {and} \bibinfo{person}{{\L}ukasz Bolikowski}.}
  \bibinfo{year}{2015}\natexlab{}.
\newblock \showarticletitle{CERMINE: automatic extraction of structured
  metadata from scientific literature}.
\newblock \bibinfo{journal}{\emph{International Journal on Document Analysis
  and Recognition (IJDAR)}} \bibinfo{volume}{18}, \bibinfo{number}{4}
  (\bibinfo{year}{2015}), \bibinfo{pages}{317--335}.
\newblock


\bibitem[\protect\citeauthoryear{Tonscheidt}{Tonscheidt}{2015}]%
        {56}
\bibfield{author}{\bibinfo{person}{Konstantin Tonscheidt}.}
  \bibinfo{year}{2015}\natexlab{}.
\newblock \showarticletitle{Leveraging code clone detection for the incremental
  migration of cloned product variants to a software product line: An
  explorative study}.
\newblock \bibinfo{journal}{\emph{Bachelorarbeit,
  Otto-von-Guericke-Universit{\"a}t Magdeburg}} (\bibinfo{year}{2015}),
  \bibinfo{pages}{4--16}.
\newblock


\bibitem[\protect\citeauthoryear{Wahlroos et~al\mbox{.}}{Wahlroos
  et~al\mbox{.}}{2019}]%
        {29}
\bibfield{author}{\bibinfo{person}{Kristian Wahlroos} {et~al\mbox{.}}}
  \bibinfo{year}{2019}\natexlab{}.
\newblock \showarticletitle{Software Plagiarism Detection Using N-grams}.
\newblock  (\bibinfo{year}{2019}).
\newblock


\bibitem[\protect\citeauthoryear{Wang, Zhang, and Yan}{Wang
  et~al\mbox{.}}{2017}]%
        {23}
\bibfield{author}{\bibinfo{person}{Ke Wang}, \bibinfo{person}{Liping Zhang},
  {and} \bibinfo{person}{Sheng Yan}.} \bibinfo{year}{2017}\natexlab{}.
\newblock \showarticletitle{A study on code clone evolution analysis}. In
  \bibinfo{booktitle}{\emph{2017 8th IEEE International Conference on Software
  Engineering and Service Science (ICSESS)}}. IEEE, \bibinfo{pages}{340--345}.
\newblock


\bibitem[\protect\citeauthoryear{Wang, Svajlenko, Wu, Xu, and Roy}{Wang
  et~al\mbox{.}}{2018}]%
        {67}
\bibfield{author}{\bibinfo{person}{Pengcheng Wang}, \bibinfo{person}{Jeffrey
  Svajlenko}, \bibinfo{person}{Yanzhao Wu}, \bibinfo{person}{Yun Xu}, {and}
  \bibinfo{person}{Chanchal~K Roy}.} \bibinfo{year}{2018}\natexlab{}.
\newblock \showarticletitle{CCAligner: a token based large-gap clone detector}.
  In \bibinfo{booktitle}{\emph{Proceedings of the 40th International Conference
  on Software Engineering}}. \bibinfo{pages}{1066--1077}.
\newblock


\bibitem[\protect\citeauthoryear{Wu, Zou, Dou, Yang, Yang, Cheng, Liang, and
  Jin}{Wu et~al\mbox{.}}{2020}]%
        {66}
\bibfield{author}{\bibinfo{person}{Yueming Wu}, \bibinfo{person}{Deqing Zou},
  \bibinfo{person}{Shihan Dou}, \bibinfo{person}{Siru Yang},
  \bibinfo{person}{Wei Yang}, \bibinfo{person}{Feng Cheng},
  \bibinfo{person}{Hong Liang}, {and} \bibinfo{person}{Hai Jin}.}
  \bibinfo{year}{2020}\natexlab{}.
\newblock \showarticletitle{SCDetector: software functional clone detection
  based on semantic tokens analysis}. In \bibinfo{booktitle}{\emph{Proceedings
  of the 35th IEEE/ACM International Conference on Automated Software
  Engineering}}. \bibinfo{pages}{821--833}.
\newblock


\bibitem[\protect\citeauthoryear{Zhang, Hall, and Baddoo}{Zhang
  et~al\mbox{.}}{2011}]%
        {36}
\bibfield{author}{\bibinfo{person}{Min Zhang}, \bibinfo{person}{Tracy Hall},
  {and} \bibinfo{person}{Nathan Baddoo}.} \bibinfo{year}{2011}\natexlab{}.
\newblock \showarticletitle{Code bad smells: a review of current knowledge}.
\newblock \bibinfo{journal}{\emph{Journal of Software Maintenance and
  Evolution: research and practice}} \bibinfo{volume}{23}, \bibinfo{number}{3}
  (\bibinfo{year}{2011}), \bibinfo{pages}{179--202}.
\newblock


\bibitem[\protect\citeauthoryear{Zibran and Roy}{Zibran and Roy}{2012}]%
        {18}
\bibfield{author}{\bibinfo{person}{Minhaz~F Zibran} {and}
  \bibinfo{person}{Chanchal~K Roy}.} \bibinfo{year}{2012}\natexlab{}.
\newblock \showarticletitle{The road to software clone management: A survey}.
\newblock \bibinfo{journal}{\emph{Dept. Comput. Sci., Univ. of Saskatchewan,
  Saskatoon, SK, Tech. Rep}}  \bibinfo{volume}{3} (\bibinfo{year}{2012}).
\newblock


\bibitem[\protect\citeauthoryear{Zimmermann}{Zimmermann}{2016}]%
        {48}
\bibfield{author}{\bibinfo{person}{Thomas Zimmermann}.}
  \bibinfo{year}{2016}\natexlab{}.
\newblock \showarticletitle{Card-sorting: From text to themes}.
\newblock In \bibinfo{booktitle}{\emph{Perspectives on data science for
  software engineering}}. \bibinfo{publisher}{Elsevier},
  \bibinfo{pages}{137--141}.
\newblock


\end{thebibliography}

\appendix

\end{document}